\documentclass{elsart}
\usepackage{amsmath}
\usepackage{amssymb}
 \newcommand{\beq}{\begin{equation}}
\newcommand{\eeq}{\end{equation}} 
\newcommand{\bea}{\begin{eqnarray}}
\newcommand{\eea}{\end{eqnarray}}

 \newcommand{\qm}{quantum
mechanics} 
\newcommand{\ca}{$C^*$-algebra} 
 \newcommand{\rep}{representation}
\newcommand{\irrep}{irreducible representation}
\newcommand{\Hs}{Hilbert space}

\newcommand{\id}{\mbox{\rm id}}

 \newcommand{\ovl}{\overline}
 \newcommand{\til}{\tilde}
\newcommand{\raw}{\rightarrow} 
\newcommand{\rac}{\circlearrowleft}
\newcommand{\lac}{\circlearrowright}
\newcommand{\law}{\leftarrow} 
\newcommand{\hraw}{\hookrightarrow} 
\newcommand{\lraw}{\leftrightarrow}

\newcommand{\ot}{\otimes}

\newcommand{\rst}{\upharpoonright} 
\newcommand{\x}{\times} 
 
\newcommand{\Co}{{\rm Co}}

\newcommand{\cin}{C^{\infty}} \newcommand{\cci}{C^{\infty}_c}

\newcommand{\inv}{^{-1}}

\newcommand{\er}{\eqref}
\newcommand{\al}{\alpha} \newcommand{\bt}{\beta}
\newcommand{\gm}{\gamma} \newcommand{\Gm}{\Gamma}
\newcommand{\dl}{\delta} \newcommand{\Dl}{\Delta}

\newcommand{\rh}{\rho} \newcommand{\sg}{\sigma}
 \newcommand{\ta}{\tau} 
 \newcommand{\phv}{\varphi}
\newcommand{\ch}{\chi} \newcommand{\ps}{\psi} 
\newcommand{\om}{\omega} \newcommand{\Om}{\Omega}

 \newcommand{\GG}{\mathfrak{G}}

\newcommand{\GS}{\mathfrak{S}} \newcommand{\g}{\mathfrak{g}}
\newcommand{\h}{\mathfrak{h}} 
 
\newcommand{\GP}{\mathfrak{P}} 

 \newcommand{\CF}{{\mathcal F}}
\newcommand{\CE}{{\mathcal E}}
 \renewcommand{\H}{{\mathcal H}}
\newcommand{\CJ}{{\mathcal J}}

 \newcommand{\CS}{{\mathcal S}}
\newcommand{\CO}{{\mathcal O}} 
\newcommand{\CQ}{{\mathcal Q}}

\newcommand{\C}{{\mathbb C}} 
 
 \newcommand{\R}{{\mathbb R}}
\newcommand{\T}{{\mathbb T}} \newcommand{\Z}{{\mathbb Z}}
 \newcommand{\SG}{\mathsf{G}}

  \makeatletter
\newskip\tempskip \def\endproof{{\parfillskip24\p@ plus\@ne
fil\@@par}\tempskip\prevdepth
\ifdim\lastskip=\z@\tempskip\z@\else\vskip-\lastskip
\ifdim\tempskip>4\p@ \tempskip.5\tempskip \else \tempskip\z@\fi\fi
\nobreak\vskip-\baselineskip\vskip-\tempskip\noindent\hbox
to\hsize{\hfill
$\blacksquare$}\par\vskip\tempskip\vskip\abovedisplayskip\@doendpe}
\makeatother \makeatletter
\newskip\tempskip \def\endiproof{{\parfillskip24\p@ plus\@ne
fil\@@par}\tempskip\prevdepth
\ifdim\lastskip=\z@\tempskip\z@\else\vskip-\lastskip
\ifdim\tempskip>4\p@ \tempskip.5\tempskip \else \tempskip\z@\fi\fi
\nobreak\vskip-\baselineskip\vskip-\tempskip\noindent\hbox
to\hsize{\hfill
$\Box$}\par\vskip\tempskip\vskip\abovedisplayskip\@doendpe}
\makeatother 

 \newtheorem{Definition}{Definition}
\newtheorem{Lemma}{Lemma}
\newtheorem{Theorem}{Theorem}
\newtheorem{Proposition}{Proposition}

\newtheorem{Corollary}{Corollary}
\newcommand{\La}{Lie algebroid}
\newcommand{\Lg}{Lie groupoid}
\newcommand{\ncg}{noncommutative geometry}
\newcommand{\PM}{Poisson manifold}
\newcommand{\KK}{\mathfrak{KK}}
\newcommand{\spinc}{\mathrm{Spin}^c}
\def\Dslash{\setbox0=\hbox{$D$}D\hskip-\wd0\hbox to\wd0{\hss\sl/\/\hss}}
\newcommand{\DS}{\Dslash}
\begin{document}
\begin{frontmatter}
\title{Lie Groupoids and Lie algebroids in physics and noncommutative geometry}
\author{N.P. Landsman}
\address{Institute for Mathematics, Astrophysics, and Particle Physics\\
Radboud University Nijmegen\\
Postbus 9010\\      6500 GL NIJMEGEN  \\
THE NETHERLANDS \\ \mbox{} \hfill \\ email: \texttt{landsman@math.ru.nl}}
\begin{abstract}
Groupoids generalize groups, spaces, group actions, and equivalence relations. This last aspect dominates in \ncg, where groupoids  provide the basic tool to desingularize pathological quotient spaces.  
In physics, however,  the main role of groupoids is to provide a unified description of internal and external symmetries. What is shared by \ncg\ and physics is the importance of Connes's idea of associating a \ca\ $C^*(\Gm)$
to a Lie groupoid $\Gm$: in \ncg\ $C^*(\Gm)$ replaces a given singular quotient space by an appropriate noncommutative space, whereas in physics it gives the algebra of observables of a quantum system whose symmetries are encoded by $\Gm$. Moreover, Connes's map $\Gm\mapsto C^*(\Gm)$ has a classical analogue $\Gm\mapsto A^*(\Gm)$ in symplectic geometry due to Weinstein, which defines the \PM\ of the corresponding classical system as the dual of the so-called \La\  $A(\Gm)$ of the \Lg\ $\Gm$, an object generalizing both Lie algebras and tangent bundles. 

Only a handful of physicists appear to be familiar with \Lg s and \La s, whereas  the  latter are practically unknown even to mathematicians working in \ncg: so much the worse for its  relationship with symplectic geometry!
Thus the aim of this review paper is to explain the relevance of both objects to both audiences.  We do so by outlining the road from canonical quantization to \Lg s and \La s via Mackey's imprimitivity theorem and its symplectic counterpart.
This will also lead the reader into symplectic groupoids, which define a `classical' category on which quantization may speculatively be defined as a functor into the category $\mathfrak{KK}$ defined by Kasparov's bivariant K-theory of \ca s. This functor unifies deformation quantization and geometric quantization,  the conjectural functoriality of quantization counting the ``quantization commutes with reduction" conjecture of Guillemin and Sternberg among its many consequences. 
\end{abstract}

\end{frontmatter}
\tableofcontents
\section{Introduction}\label{intro}
 Influenced by mathematicians such as Grothendieck,  Mackey,  Connes, and Weinstein, the use of groupoids in pure mathematics has  become respectable (though by no means widespread), at least  in their respective areas of  algebraic geometry, representation theory, noncommutative geometry, and symplectic geometry.\footnote{There is a Groupoid Home Page at \texttt{http://unr.edu/homepage/ramazan/groupoid/}.
See also \texttt{http://www.cameron.edu/$\sim$koty/groupoids/} for an incomplete but useful list of papers involving groupoids, necessarily restricted to mathematics.}
 Unfortunately, in physics groupoids remain virtually unknown.\footnote{Conferences such as {\it Groupoids in Analysis, Geometry, and Physics} (Boulder, 1999, see \cite{RR}) and {\it Groupoids and Stacks in Physics and Geometry} (Luminy, 2004) tend te be almost exlusively attended by mathematicians.} 
This is a pity for at least two reasons. Firstly, much of the spectacular mathematics developed in the areas just mentioned becomes inaccessible to physicists, despite its undeniable relevance to physics. This obstructs, for example, the development of  a good theory for quantizing singular spaces (of the kind necessary for quantum cosmology); cf.\ \cite{LPS}. As a case in point, many completely natural constructions in \ncg\ look mysterious to physicists who are not familiar with groupoids. 
Secondly, in the smooth setting, Lie groupoids along with their associated infinitesimal objects called Lie algebroids provide an ideal framework for practically all aspects of both classical and quantum physics that involve symmetry in one way or the other. 

Indeed, whereas in the work of Grothendieck and Connes groupoids mainly occur as generalizations of equivalence relations,\footnote{Grothendieck (to R. Brown in a letter from 1985): {\it ``The idea of making systematic use of groupoids (notably fundamental groupoids of spaces, based on a given set of base points), however evident as it may look today, is to be seen as a significant conceptual advance, which has spread into the most manifold areas of mathematics. (\ldots) In my own work in algebraic geometry, I have made extensive use of groupoids - the first one being the theory of the passage to quotient by a `pre-equivalence relation' (which may be viewed as being no more, no less than a groupoid in the category one is working in, the category of schemes say), which at once led me to the notion (nowadays quite popular) of the nerve of a category. The last time has been in my work on the Teichm\"{u}ller tower, where working with a `Teichm\"{u}ller groupoid' (rather than a `Teichm\"{u}ller group') is a must, and part of the very crux of the matter (\ldots)''}} the role of groupoids as generalized symmetries has been emphasized by Weinstein \cite{Wei2}: 
 {\it ``Mathematicians tend to think of the notion of symmetry as being virtually synonymous with the theory of}  groups {\it and their actions.\footnote{Cf.\ Connes: {\it ``It is fashionable among mathematicians to despise groupoids and to consider that only groups have an authentic mathematical status, probably because of the pejorative suffix oid.''} \cite{Con}}
 (\ldots) In fact, though groups are indeed sufficient to characterize homogeneous structures, there are plenty of objects which exhibit what we clearly recognize as symmetry, but which admit few or no nontrivial automorphisms. It turns out that the symmetry, and hence much of the structure, of such objects can be characterized if we use}  groupoids {\it and not just groups.}  

The aim of this paper is to (briefly) explain what \Lg s and \La s are, and 
(more extensively) to outline which role they play in physics (at least from the perspective of the author). Because of the close relationship between quantum theory and \ncg\ on the one hand, and classical mechanics and symplectic geometry on the other,\footnote{Throughout this paper we use the term `symplectic geometry' so as to include Poisson geometry.} our discussion obviously relates to matters of pure mathematics as well, and here the physics perspective turns out to be quite useful in clarifying the relationship between noncommutative and symplectic geometry. This relationship is rarely studied in \ncg, which might explain the regrettable absence of the concept of a \La\ from the field.\footnote{Except for the work of the author, the sole exception known to him is \cite{NT}.} 

With this goal in mind, one of our main points will be to show that the role of \Lg s on the quantum or noncommutative side is largely paralleled by the role \La s play on the classical or symplectic side. The highlight of this philosophy is undoubtedly  the close analogy between Connes's map $\Gm\mapsto C^*(\Gm)$ in \ncg\ \cite{Con} and Weinstein's map $\Gm\mapsto A^*(\Gm)$ in symplectic geometry \cite{CDW,Cou}, notably the functoriality of both \cite{Lan2}. Furthermore, the transition from classical to quantum theory through deformation quantization turns out to be given precisely by the association of the \ca\ $C^*(\Gm)$ to the \PM\ $A^*(\Gm)$ \cite{Lan1,LR,Ram}. Hence quantization is closely related to `integration', in the sense of  the association of a Lie groupoid to a \La; see \cite{Mackenzie95} for an introduction to this problem, and  \cite{CF1} for its solution. 

 We do not provide an extensive mathematical introduction to \Lg s and \La s, 
 partly because we have already done so before \cite{Lan}, and partly 
 because various excellent textbooks on this subject are now available  \cite{McK,MM,Weinsteinplus}. Instead, we start entirely on the physics side, with a crash course on canonical quantization and its reformulation by Mackey in terms of systems of imprimitivity. In its original setting Mackey's notion of quantization was not only limited to homogeneous configuration spaces, but in addition lacked an underlying classical theory.\footnote{More precisely, the underlying classical theory was not correctly identified \cite{Mac98}.} Both drawbacks are entirely removed once one adopts the perspective of \Lg s on the quantum side and \La s on the classical side, and we propose this as a convenient point of entry for physicists into the world of these seemingly strange and unfamiliar objects. 
 
 Once this perspective has been adopted, the entire theory of canonical quantization and its (finite-dimensional) generalizations is absorbed into a single theorem, stating that the association of $C^*(\Gm)$ to $A^*(\Gm)$
 mentioned above is a `strict' deformation quantization (in the sense of Rieffel \cite{Rie89,Rie94}). 
 Furthermore, in our opinion the deepest understanding of Mackey's imprimitivity theorem comes from its derivation from the functoriality
 of Connes's map $\Gm\mapsto C^*(\Gm)$; similarly, the classical analogue of the imprimitivity theorem in symplectic geometry \cite{Ziegler} can be derived from the  functoriality of Weinstein's map $\Gm\mapsto A^*(\Gm)$ already mentioned. 

 We finally combine  the toolkit of \ncg\ with that of symplectic geometry in
proposing a functorial approach to quantization, which is based on KK-theory on the quantum side and on symplectic groupoids on the classical side. As we see it, this approach provides the ultimate generalization of the `quantization commutes with reduction' philosophy of Dirac \cite{Dir64} (in physics) and Guillemin and Sternberg \cite{GS,GGK} (in mathematics). Beside the use of the K-theory of \ca s,  this generalization hinges on the use of \Lg s and \La s, and therefore appears to be an appropriate  endpoint of this paper.  
\section{From canonical quantization to systems of imprimitivity}
Quantum mechanics was born in 1925 with the work of Heisenberg, who discovered  the noncommutative structure of its algebra of observables \cite{Heis}. The complementary work of Schr\"{o}dinger from 1926 \cite{Wave}, on the other hand, rather started from the classical geometric structure of configuration space. Within a year, their work was unified by von Neumann, who introduced the abstract concept of a Hilbert space, in which 
Schr\"{o}dinger's wave functions are vectors, and Heisenberg's observables are linear operators; see \cite{JvN}. As every physicist knows, the basic link between  matrix mechanics and  wave mechanics lies in the identification of
Heisenberg's infinite matrices $p_j$ and $q^i$ ($i,j=1,2,3$), representing the momentum and position of a particle moving in $\R^3$, with Schr\"{o}dinger's operators $-i\hbar\partial/\partial x^j$ and  $x^i$ (seen as a multiplication operator) on the \Hs\ $\H=L^2(\R^3)$, respectively. The key to this identification lies in the canonical commutation relations
\beq [p_i,q^j]=-i\hbar\dl^j_i. \label{ccr}\eeq
Although a mathematically rigorous theory of these commutation relations (as they stand) exists \cite{JoMo,Schmuedgen}, they are problematic nonetheless. Firstly, the operators involved are unbounded, and in order to represent physical observables they have to be self-adjoint; yet on their respective domains of self-adjointness the commutator on the left-hand side is undefined. Secondly, \er{ccr} relies on the possibility of choosing  global coordinates on $\R^3$, which precludes at least a naive generalization to arbitrary configuration spaces.\footnote{Mackey \cite[p.\ 283]{Mac92}: {\it ``Simple and elegant as this model is, it appears at first sight to be quite arbitrary and} ad hoc. {\it It is difficult to understand how anyone could have guessed it and by no means obvious how to modify it to fit a model for space different from $\R^r$.}} 

Finding an appropriate mathematical interpretation of the canonical commutation relations \er{ccr} is the subject of quantization theory; see \cite{AE,Lan} for recent reviews. From the numerous ways to handle the situation, we here select Mackey's approach \cite{Mac68,Mac92}.\footnote{Continuing the previous quote, Mackey claims with some justification that his approach {\it ``(a) Removes much of the mystery. (b) Generalizes in a straightforward way to any model for space with a separable locally compact group of isometries. (c) Relates in an extremely intimate way to [the theory of induced \rep s].''} 
In any case, Mackey's  approach to the  canonical commutation relations, especially in its \ca ic reformulation presented below, is vastly superior to their equally \ca ic reformulation in terms of the so-called Weyl $C^*$-algebra (cf.\ e.g.\ \cite{BR2}). Indeed (see \cite{Lan} Def.\ IV.3.5.1), the Weyl algebra over a \Hs\ $\H$ (which in the case at hand is $\C^3$)
may be seen as the twisted group \ca\ over $\H$ as an abelian group under addition, {\it equipped with the discrete topology}. This rape of $\H$ as a topological space is  so ugly that it is surprising that papers on the Weyl $C^*$-algebra continue to appear.
Historically, Weyl's exponentiation of the canonical commutation relations was just one of the first attempts to reformulate a problem involving unbounded operators in terms of bounded ones, and has now been superseded.} The essential point is to assign momentum and position a quite different role in \qm, despite the fact that in classical mechanics $p$ and $q$ can be interchanged by a canonical transformation.\footnote{This feature is shared by most approaches to quantization, except the one mentioned in the preceding footnote.} 

Firstly, the position operators $q^j$ are collectively replaced by a single projection-valued measure $P$ on $\R^3$,\footnote{A projection-valued measure $P$ on a space $\Om$ with Borel structure (i.e.\ equipped with a $\sg$-algebra of measurable sets defined by the topology) with values in a \Hs\ $\H$ is a map $E\mapsto P_E$ from the Borel subsets
$E\subset\Om$ to the projections on $H$ that satisfies $P_{\emptyset}=0$,
$P_{\Om}=1$, $P_E P_F=P_FP_E=P_{E\cap F}$ for all measurable $E,F\subset\Om$, and
$P_{\cup_{i=1}^{\infty} E_i}=\sum_{i=1}^{\infty} P_{E_i}$ for all countable collections of mutually disjoint $E_i\subset\Om$.} which is given by $P_E=\ch_E$ as a multiplication operator on $L^2(\R^3)$. Given this $P$, any multiplication operator $f$ defined by a measurable function $f:\R^3\raw\R$ can be represented as $f=\int_{\R^3} dP_E(x)\, f(x)$,
which is defined and self-adjoint on a suitable domain.\footnote{This domain consists of all
$\ps\in H$ for which $\int_{\R^3} d(\ps,P_E(x)\ps)\, |f(x)|^2<\infty$.} In particular, the position operators $q^i$ can be reconstructed from $P$ by choosing $f(x)=x^i$.

Secondly,  the momentum operators $p_i$ are collectively replaced by a single unitary group \rep\ $U(\R^3)$ on $L^2(\R^3)$, defined by $$U(y)\ps(x):=\ps(x-y).$$ Each $p_i$
can be reconstructed from $U$ by means of $$p_i\ps:=i\hbar \lim_{t_i\raw 0} t_i\inv(U(t_i)-1)\ps,$$ where $U(t_i)$ is $U$ at $x^i=t_i$ and $x^j=0$ for $j\neq i$; this operator is defined and self-adjoint on the set of all $\ps\in H$ for which the limit exists (Stone's theorem \cite{Ped2}). 

 Consequently, it entails no loss of generality to work with the pair $(P,U)$ instead of the   $(q^j,p_i)$. The commutation relations \er{ccr} are now replaced by 
\beq U(x)P_EU(x)\inv=P_{xE}, \label{impr}\eeq
where $E$ is a Borel subset of $\R^3$ and $xE=\{x\om\mid\om\in E\}$. On the basis of this reformulation, Mackey proposed  the following sweeping generalization of the the canonical commutation relations.\footnote{In order to maintain the connection with the classical theory later on, we restrict ourselves
to Lie groups acting smoothly on manifolds. Mackey actually formulated his results more generally in terms of separable locally compact groups acting continuously on locally compact spaces.}
\begin{Definition}\label{soi}
Suppose a Lie group $G$ acts smoothly on a manifold $M$. 
\begin{enumerate}
\item 
A {\rm system of imprimitivity} $(\H,U,P)$ for this action consists of a \Hs\ $\H$, a unitary \rep\ $U$ of $G$ on $\H$, and a projection-valued measure $E\mapsto P_E$
on $M$ with values in $\H$, such that  \er{impr} holds for all $x\in G$ and all Borel sets $E\subset M$.
\item A  {\rm $G$-covariant \rep} $(\H,U,\pi)$ of the \ca\ $C_0(M)$  relative to this action consists of a \Hs\ $\H$, a unitary \rep\ $U$ of $G$ on $\H$, and  a nondegenerate \rep\ $\pi$ of $C_0(M)$ on $\H$ satisfying 
\beq U(x)\pi(\phv)U(x)\inv=\pi(L_x\phv) \label{Gcov}\eeq
 for all $x\in G$ and $\phv\in C_0(M)$, where $L_x\phv (m)=\phv(x\inv m)$.
\end{enumerate}
 \end{Definition}
 The spectral theorem (cf.\ \cite{Ped2}) implies that these notions are equivalent: a projection-valued measure $P$ defines and is defined by a nondegenerate \rep\ $\pi$ of $C_0(M)$ on $\H$ by means of $\pi(\phv)=\int_M dP(m)\, \phv(m)$, and \er{impr} is then equivalent to the covariance condition \er{Gcov}. Hence we may interchangeably speak of systems of imprimitivity or covariant \rep s.
  As a further reformulation, it is
 easy to show (cf.\ \cite{DKR,EH,Ped1}) that there is a bijective correspondence between $G$-covariant \rep s of $C_0(M)$ and nondegenerate \rep s of the 
 so-called transformation group \ca\ $C^*(G,M)\equiv G\x_{\al} C_0(M)$
 defined by the given $G$-action on $M$, which determines an automorphic action $\al$ of $G$ on $C_0(M)$ by $\al_x=L_x$.\footnote{In one direction, this correspondence is as follows: given a $G$-covariant \rep\ $(\H,\pi,U)$, one defines a \rep\ $\pi_U(C^*(G,M))$ by extension  of $\pi_U(f)=\int_G dx\, \pi(f(x,\cdot))U(x)$, where $f\in\cci(G\x M)\subset C^*(G,M)$, and $f(x,\cdot)$ is seen as an element of $C_0(M)$.} 
 
Such a system describes the \qm\ of a particle moving on a configuration space $M$ on which $G$ acts by symmetry transformations; in particular, each element $X$ of the Lie algebra $\g$ of $G$ defines a generalized momentum operator \beq
\hat{X}=i\hbar dU(X)\label{mom}\eeq on $\H$, which is defined and self-adjoint on the domain of vectors $\ps\in \H$ for which $$dU(X)\ps:=\lim_{t\raw 0} t\inv(U(\exp(tX))-1)\ps$$ exists. These operators satisfy 
the generalized canonical commutation relations
\beq [\hat{X},\hat{Y}] = i\hbar \widehat{[X,Y]}\eeq and
\beq [\hat{X},\pi(\phv)] = \pi(\xi_X\phv),\eeq
where $\phv\in\cci(M)$ and $\xi_X$ is the canonical  vector field on $M$ defined by the $G$-action; of course, these should be supplemented with \beq [\pi(\phv_1),\pi(\phv_2)]=0. \eeq
Elementary \qm\ on $\R^n$ then corresponds to the special case $M=\R^n$
and $G=\R^n$ with the usual additive group structure. 
\section{The imprimitivity theorem}
In the spirit of the \ca ic approach to quantum physics \cite{Pri,Thi,Emch,Haa}, the \ca\ $C^*(G,M)$ defined by the given  $G$-action on $M$ should be seen as an algebra of observables, whose inequivalent \irrep s define the possible superselection sectors of the system. As we have seen, these \rep s may equivalently be seen as systems of imprimitivity or as $G$-covariant \rep s of $C_0(M)$ \cite{DKR,EH,Ped1}.
In any case,  it is of some interest to classify these. 
Mackey's {\it imprimitivity theorem} describes the simplest case where this is possible. 
\begin{Theorem}\label{imthm} {\rm \cite{Bott57,Mac52}}
Let $H$ be a closed subgroup of $G$ and let $G$ act on $M=G/H$ by left translation.  Up to unitary equivalence, there is a bijective correspondence between systems of imprimitivity  $(\H,U,P)$ for this action  (or, equivalently,
$G$-covariant \rep\ of $C_0(G/H)$ or nondegenerate \rep s of the transformation group \ca\ $C^*(G,G/H)$) 
and unitary \rep s $U_{\ch}$ of $H$, as follows:
\begin{itemize}
\item 
 Given $U_{\ch}(H)$ on a \Hs\ $\H_{\ch}$, the triple $(\H^{\ch},U^{\ch},P^{\ch})$ is a system of imprimitivity, where 
$\H^{\ch}=L^2(G/H,G\x_H \H_{\ch})$ is the Hilbert space of $L^2$-sections of the vector bundle 
$G\x_H \H_{\ch}$ associated to the principal $H$-bundle $G$ over $G/H$ by $U_{\ch}$,
$U^{\ch}$ is the \rep\ of $G$ induced by $U_{\ch}$, and $P^{\ch}_E=\ch_E$ acts canonically on $\H^{\ch}$ as a  multiplication operator. 
\item Conversely, 
if $(\H,U,P)$ is a system of imprimitivity, then there exists a unitary \rep\ $U_{\ch}(H)$ such that the triple $(\H,U,P)$ is unitarily equivalent to the triple $(\H^{\ch},U^{\ch},P^{\ch})$ just described.
\end{itemize}
 The correspondence $(\H_{\ch},U_{\ch})\lraw (\H^{\ch},U^{\ch},P^{\ch})$ preserves direct sums and, accordingly,  irreducibility.
\end{Theorem}

The simplest and at the same time most beautiful application of the imprimitivity theorem is Mackey's recovery of the Stone--von Neumann uniqueness theorem concerning the (regular) irreducible \rep s of the canonical commutation relations: taking $G=\R^3$ and $H=\{e\}$ (so that $M=\R^3$), one finds that the associated system of imprimitivity possesses precisely one irreducible \rep, since the trivial group obviously has only one such \rep.\footnote{The 
``uniqueness of the canonical commutation relations'' has also been derived from  the fact that (up to unitary equivalence) there is only one \irrep\ of any of the following objects: i) The Heisenberg Lie group with given nonzero central charge (von Neumann's  theorem \cite{vNu}); ii) The Weyl \ca\ over a finite-dimensional \Hs, provided one restricts oneself to the class of regular \rep s \cite{BR2}; or iii) The \ca\ of compact operators \cite{Rie72}.}  Furthermore (and this was one of Mackey's main points), one may keep $\R^3$ as a confuguration space but replace $G=\R^3$ by the Euclidean group $G=SO(3)\ltimes\R^3$, so that $H=SO(3)$.
The generalized momenta then include the angular momentum operators $J^i$ along with their commutation relations, and the imprimitivity theorem then asserts that the \irrep s of \er{impr} correspond to the usual \irrep s $U_j$ of $SO(3)$, $j=0,1,\ldots$.\footnote{By the usual arguments, one may replace $SO(3)$ by $SU(2)$ in this argument, so as to obtain
$j=0,1/2,\ldots$.} Mackey saw this as an explanation for the emergence of spin as a purely quantum-mechanical degree of freedom; the latter perspective of spin goes back to the pionieers of quantum theory \cite{Pauli}, but is now  obsolete (see Section \ref{DQsection} below).  

Mackey's imprimitivity theorem admits a generalization to $G$-actions on an arbitrary manifold $M$, provided the action is regular.\footnote{In view of this simple result, \ca ists are mainly interested in nonregular actions, cf.\ \cite{EH}, but for physics Proposition \ref{Glimm} is quite useful.
In any case, an example of a nonregular action is the action of $\Z$ on $\T$ by irrational rotations.} 
\begin{Proposition}\label{Glimm} {\rm \cite{Gli1,Gli2}}
 Suppose that each $G$-orbit in $M$ is
(relatively) open in its closure.  The \irrep s of $C^*(G\ltimes M)$ are
classified by pairs $(\CO,U_{\ch})$, where $\CO$ is a $G$-orbit in
$M$ and $U_{\ch}$ is an \irrep\ of the stabilizer  of an arbitrary point
$m_0\in\CO$.\footnote{The associated $G$-covariant \rep\ of $C_0(M)$  may be realized  by multiplication operators on the \Hs\ $\H^{\ch}$ carrying the \rep\ $U^{\ch}(G)$ induced by $U_{\ch}$.}
\end{Proposition}
 
In view of the power of  Mackey's imprimitivity theorem, both for \rep\ theory and quantization theory, increasingly sophisticated and insightful proofs have been published over the last five decades.\footnote{Mackey's own proof was rather measure-theoretic in flavour, and did not shed much light on the origin of his result. Probably the shortest proof is \cite{Orsted79}.} All proofs relevant to \ncg\ are either based on or are equivalent to: 
\begin{Theorem}\label{RG} {\rm \cite{Green,Rie74}}
The transformation group \ca\ $C^*(G,G/H)$ is Morita equivalent to $C^*(H)$.
\end{Theorem}
This means that there exists a so-called equivalence or imprimitivity bimodule $\CE$ (which in modern terms would be called a $C^*(G,G/H)$-$C^*(H)$ Hilbert bimodule)\footnote{A Hilbert bimodule $A\lac \CE\rac B$ over \ca s $A$ and $B$ consists of a  Banach space
$\CE$ that is an algebraic $A$-$B$ bimodule, and is equipped with a
$B$-valued inner product that is compatible with the $A$ and $B$
actions. Such objects were first considered by Rieffel \cite{Rie74}, who defined 
 an `interior' tensor product $\CE \hat{\ot}_{B}\CF$
of an $A$-$B$ Hilbert bimodule $\CE$ with a $B$-$C$ Hilbert bimodule $\CF$,
which is an $A$-$C$ Hilbert bimodule.} that allows one to set up the bijective correspondence - called for in Mackey's imprimitivity theorem - between (nondegerenerate) \rep s of $C^*(G,G/H)$ and those of $C^*(H)$ (or equivalently, of $H$). Given 
a unitary \rep\ $U_{\ch}(H)$ on a \Hs\ $\H_{\ch}$, or the associated \rep\ $\pi_{\ch}$ of $C^*(H)$ on the same space, one constructs a \Hs\ 
$\H^{\ch}=\CE\hat{\otimes}_{\pi_{\ch}}\H_{\ch}$. The action of $C^*(G,G/H)$ on $\CE$ descends to an action $\pi^{\ch}(C^*(G,G/H))$ on $\H^{\ch}$, and extracting the associated \rep s of $G$ and of $C_0(G/H)$ one finds that this is precisely Mackey's induction
construction paraphrased in Theorem \ref{imthm}. Conversely, a given \rep\ $\pi^{\ch}$ of $C^*(G,G/H)$ on a \Hs\ $\H^{\ch}$ defines $\pi_{\ch}(C^*(H))$ on
$\H_{\ch}=\ovl{\CE}\hat{\otimes}_{\pi^{\ch}}\H^{\ch}$, and this process is the inverse of the previous one. 
Replacing the usual algebraic bimodule tensor product by Rieffel's 
interior tensor product $\hat{\otimes}_{\pi}$, this entirely mimics the corresponding procedure in algebra (cf.\ \cite{Faith73}); in the same spirit, one  infers also in general that two Morita equivalent  \ca s have equivalent \rep\ categories.

The reformulation of Theorem \ref{imthm} as Theorem \ref{RG} begs the question what the deeper origin of the latter could possibly be. One answer is
given by the analysis in  \cite{EKQR}, from which Theorem \ref{RG} emerges as merely a droplet in an ocean of imprimitivity theorems. The answer below \cite{Lan2,Lan3,Lan4,MRW} is equally categorical in spirit, but is entirely  based on the use of Lie groupoids. Namely, we will derive  Theorem \ref{RG} and hence Mackey's Theorem \ref{imthm} from the functoriality of Connes's map \er{Connesmap} below, which associates a \ca\ to a \Lg.
Apart from the fact that this is very much in the spirit 
of \ncg, the use of \Lg s will enable us to formulate an analogous classical procedure in terms of \La s and \PM s. All this requires a little preparation.
\section{Intermezzo: Lie groupoids}\label{intermezzo}
 Recall that  a {\it groupoid} is a small category (i.e.\ a category in which the underlying classes are sets) in which each arrow is
invertible. We denote the total space (i.e.\ the set of arrows) of a groupoid $\Gm$ by $\Gm_1$, and the base space (i.e.\ the set on which the arrows act) by $\Gm_0$; the object inclusion map $\Gm_0\hraw \Gm_1$ is written $u\mapsto 1_u$.  We denote the inverse $\Gm_1\raw \Gm_1$ by $x\mapsto x\inv$, and the source and target
maps by $s,t:\Gm_1\raw \Gm_0$. Thus the composable pairs form the space $\Gm_2:=\{(x,y)\in \Gm_1\x \Gm_1\mid s(x)=t(y)\}$, so that if $(x,y)\in \Gm_2$ then $xy\in \Gm_1$ is defined.\footnote{Thus the axioms are: 1. $s(xy)=s(y)$ and $t(xy)=t(x)$; 2. $(xy)z=x(yz)$
3. $s(1_u)=t(1_u)=u$ for all $u\in \Gm_0$; 4. $x1_{s(x)}=1_{t(x)}x=x$
for all $x\in \Gm_1$.}  A {\it Lie groupoid} is a groupoid for which $\Gm_1$
and $\Gm_0$ are manifolds ($\Gm_1$ not necessarily being Hausdorff), $s$ and $t$ are surjective submersions, and
multiplication and inversion  are smooth.\footnote{It follows that object inclusion is an immersion, that inversion  is a diffeomorphism, that $\Gm_2$ is a closed submanifold of $\Gm_1\x \Gm_1$, and that for each $u\in \Gm_0$ the fibers $s\inv(u)$ and
$t\inv(u)$ are submanifolds of $\Gm_1$.} See \cite{McK,MM} for recent textbooks on Lie groupoids and related matters.\footnote{The concept of a \Lg\ was introduced by Ehresmann.}

Some examples of Lie groupoids that are useful to  keep in mind are:
\begin{itemize} 
\item A {\it Lie group}
$G$, where $\Gm_1=G$ and $\Gm_0=\{e\}$).
\item A {\it manifold} $M$, where $\Gm_1=\Gm_0=M$ with the
obvious trivial groupoid structure $s(x)=t(x)=1_x=x\inv = x$, and $xx=x$.
\item The {\it pair groupoid} over a manifold $M$, where $\Gm_1=M\x M$ and
$\Gm_0=M$, with  $s(x,y)=y$, $t(x,y)=x$, $(x,y)\inv=(y,x)$, 
$(x,y)(y,z)=(x,z)$, and $1_x=(x,x)$.
\item The {\it gauge groupoid} defined by a principal $H$-bundle $P\stackrel{\pi}{\raw}M$,
where $\Gm_1=P\x_H P$ (which stands for $(P\x P)/H$ with respect to the diagonal $H$-action on $P\x P$), $\Gm_0=M$, $s([p,q])=\pi(q)$, $t([p,q])=\pi(p)$, $[x,y]\inv=[y,x]$, and $[p,q][q,r]=[p,r]$ (here $[p,q][q',r]$
is defined whenever $\pi(q)=\pi(q')$, but to write down the product one picks $q\in\pi\inv(q')$). 
\item The {\it action groupoid} $G\ltimes M$ defined by a smooth (left) action $G\lac M$ of a Lie group $G$ on a manifold $M$, where  $\Gm_1=G\x M$, $\Gm_0=M$,
$s(g,m)=g\inv m$, $t(g,m)=m$, $(g,m)\inv=(g\inv,g\inv m)$, and $(g,m)(h,g\inv m)=(gh,m)$.
\end{itemize}
  
As mentioned before, an equivalence relation on a set $M$ defines a groupoid, namely the 
obvious subgroupoid of the pair groupoid over $M$. However, in interesting examples this is rarely a Lie groupoid. To obtain a Lie groupoid resembling a given equivalence relation on a manifold, various refinements of the subgroupoid in question have been invented, of which the holonomy groupoid defined by a foliation is
the most important example  for \ncg\ \cite{Con,MM,Phi}. 

For reasons to emerge from the ensuing story, we look  at Lie groupoids as objects in the {\it category of principal bibundles}. To define this category, we first recall that an action of a groupoid $\Gm$ on a space $M$ is only defined if $M$ comes equipped with a map $M\stackrel{\pi}{\raw} \Gm_0$. In that case,  a left $\Gm$ action on $M$ is a 
map $(x,m)\mapsto xm$ from $\Gm_1 \times^{s,\pi}_{\Gm_0}M$ to $M$,\footnote{Here we use the notation $A\times^{f,g}_B C= \{(a,c)\in A\x C\mid f(a)=g(c)\}$
for the fiber product
of sets $A$ and $C$ with respect to maps $f:A\raw B$ and $g:C\raw B$.}
 such that $\pi(xm)=t(x)$, $xm=m$ for all $x\in \Gm_0$, and $x(ym)=(xy)m$
whenever $s(y)=\ta(m)$ and $t(y)=s(x)$.  Similarly, given a map $M\stackrel{\rh}{\raw} \Delta_0$, a right action of a
groupoid $\Delta$ on $M$ is a map $(m,h)\mapsto mh$ from
$M\times^{\rh,t}_{\Delta_0} \Delta_1$ to $M$ that satisfies $\rh(mh)=s(h)$, $mh=m$
for all $h\in \Delta_0$, and $(mh)k=m(hk)$ whenever $\rh(m)=t(h)$ and
$t(k)=s(h)$. Now,  if $\Gm$ and $\Delta$ are groupoids, a
$\Gm$-$\Delta$ bibundle $M$, also written as $\Gm\lac M\rac \Delta$, carries a left $\Gm$ action as well as a right $\Delta$-action that commute.\footnote{That is,
 one has $\ta(mh)=\ta(m)$, $\rh(xm)=\rh(m)$, and $(xm)h=x(mh)$ 
whenever defined.}  
Such a bibundle is called {\it principal} when $\pi:M\raw \Gm_0$ is 
surjective, and the $\Delta$
action is free (in that $mh=m$ iff $h\in \Delta_0$) and transitive along
the fibers of $\pi$. 

Suppose one has right principal bibundles $\Gm\lac M\rac \Delta$ and $\Delta\lac N\rac \Theta$.
The fiber product $M\times_{\Delta_0} N$ 
carries a right $\Delta$ action, given by
$h:(m,n)\mapsto (mh,h\inv n)$ (defined as appropriate). The orbit space $(M\times_{\Delta_0} N)/\Delta$ 
 is a $\Gm$-$\Theta$ bibundle in the obvious way inherited from the original actions. Thus, regarding $\Gm\lac M\rac \Delta$ as an arrow from $\Gm$ to $\Delta$ and $\Delta\lac N\rac \Theta$ as an arrow from $\Delta$ to $\Theta$, one map look upon $\Gm\lac (M\times_\Delta N)/\Delta\rac \Theta$ as an arrow from $\Gm$ to $\Theta$, defining the product or composition of $M$ and $N$. However, this product is associative merely up to isomorphism, so that in order to have a category one should regard isomorphism classes of principal bibundles as arrows. 

For Lie groupoids everything in these definitions has to be smooth  (and $\pi$ a surjective submersion).
\begin{Definition} \label{CatG}{\rm \cite{C82,Hae,HS,Moe,MM}}
The category $\mathfrak{G}$ of Lie groupoids and principal bibundles has Lie groupoids as objects and isomorphism classes $[\Gm\lac M\rac \Delta]$ of principal bibundles as arrows. Composition of arrows is given by $$
[\Gm\lac M\rac \Delta]\circ [\Delta\lac N\rac \Theta]=[\Gm\lac (M\times_\Delta N)/\Delta\rac \Theta],$$
and the identities are given by $1_{\Gm}= [\Gm\lac\Gm\rac\Gm]$, seen as a bibundle in the obvious way.
\end{Definition}

Of course, it can be checked that this definition is correct in the sense that one indeed defines a category in this way. This category has the remarkable feature that (Morita) equivalence of groupoids (as defined in \cite{MRW}, a notion heavily used in \ncg) is the same as isomorphism of objects in $\mathfrak{G}$. 
\section{From  Lie groupoids to the imprimitivity theorem}
A central idea in \ncg\ is the association 
\beq \Gm\mapsto C^*(\Gm) \label{Connesmap}\eeq
of a \ca\ $C^*(\Gm)$ to a Lie groupoid $\Gm$ \cite{Con}.\footnote{See also \cite{Lan,LR,Pat} for  detailed presentations.  For a Lie groupoid $\Gm$ Connes's $C^*(\Gm)$ is the same (up to isomorphism of \ca s) as the \ca\ Renault associates to a locally compact groupoid with Haar system \cite{Ren}, provided one takes the Haar system canonically defined by the smooth structure on $\Gm$.}
Here $C^*(\Gm)$ is a suitable completion of the function space $\cci(\Gm_1)$, equipped with a convolution-type product defined by the groupoid structue. For the above examples, this yields:
\begin{itemize} 
\item The \ca\ of a Lie group $G$ is the usual convolution \ca\ $C^*(G)$ defined by the Haar measure on $G$ \cite{Ped1}.
\item For a manifold $M$ one has $C^*(M)=C_0(M)$.
\item The pair groupoid over a connected manifold $M$ defines $C^*(M\x M)\cong K(L^2(M))$, i.e.\ the \ca\ of compact operators on the $L^2$-space canonically defined by a manifold.
\item The  \ca\ defined by a gauge groupoid $P\times_H P$ as above is isomorphic to $K(L^2(M))\otimes C^*(H)$
(but any explicit isomorphism depends on the choice of a measurable section $s:M\raw P$, which in general cannot be smooth). 
\item For an action groupoid defined by $G\lac M$ one has $C^*(G\ltimes M)\cong C^*(G,M)$, the transformation group \ca\ defined by the given action \cite{EH,Ped1}.
\end{itemize}

Having already defined the category $\GG$ of principal bibundles for \Lg s, in order to make the map  \er{Connesmap} functorial, one  has to regard \ca s as objects in a suitable category $\mathfrak{C}$ as well. 
\begin{Definition} \label{CatC} {\rm \cite{EKQR,Lan3,Schweizer}} 
The category $\mathfrak{C}$ has \ca s as objects and 
 isomorphism classes $[A\lac \CE\rac B]$ of Hilbert bimodules, as arrows, 
composed using Rieffel's interior tensor product.  The identities are given by $1_A=A\lac A\rac A$, defined in the obvious way. 
\end{Definition}

A crucial feature of this construction is that the notion of isomorphism of objects in $\mathfrak{C}$ coincides with Rieffel's (strong) Morita equivalence of \ca s.
\begin{Theorem} \label{Lan2Thm}{\rm \cite{Lan2}}
Connes's map
 $\Gm\mapsto C^*(\Gm)$ is functorial from  the category $\mathfrak{G}$
 of Lie groupoids and principal bibundles  to the category  $\mathfrak{C}$
of \ca s and Hilbert bimodules.
\end{Theorem}
\begin{Corollary} \label{Lan4cor} {\rm \cite{Lan4,MRW}}
 Connes's map  $\Gm\mapsto C^*(\Gm)$ preserves Morita equivalence,
in the sense that if $\Gm$ and $\Dl$ are Morita equivalent Lie groupoids, then $C^*(\Gm)$ and $C^*(\Dl)$ are Morita equivalent \ca s.\end{Corollary}

 The imprimitivity bimodule $C^*(\Gm)\lac \mathcal{E}\rac C^*(\Dl)$ establishing the Morita equivalence of $C^*(\Gm)$ and $C^*(\Dl)$ is obtained from the principal bibundle $\Gm\lac E\rac\Dl$ establishing the Morita equivalence of $\Gm$ and $\Dl$ in a very simple way, amounting to the completion of
$\cci(\Gm)\lac\cci(E)\rac\cci(\Dl)$; see \cite{Lan2,StadlerUchi}.

For example, in Mackey's case one has $\Gm=G\ltimes (G/H)$ and $\Dl=H$, linked by the principal bibundle $G\ltimes (G/H)\lac G\rac H$ in the obvious way;\footnote{For example, $(g_1,m)g_2=g_1g_2$, defined whenever $m=\pi(g_1g_2)$.} the associated imprimitivity bimodule for $C^*(G\ltimes (G/H))\cong C^*(G,G/H)$ and $C^*(H)$ is precisely the one found by Rieffel \cite{Rie74}. Thus 
Theorem \ref{RG}, and thereby Mackey's imprimitivity theorem, ultimately derives from the Morita equivalence  
\beq G\ltimes (G/H)\sim H \label{GHGH}\eeq of groupoids, which is an almost trivial fact once the appropriate framework has been set up. This framework cannot be specified in terms of groups and group actions alone, despite the fact that the two groupoids relevant to Mackey's imprimitivity theorem reduce to those.

 Mackey's analysis of the canonical commutation relations admits various other generalizations than Proposition \ref{Glimm}, at least one of which is related to groupoids as well: instead of 
generalizing the action groupoid $G\ltimes (G/H)$ to an arbitrary action groupoid
$G\ltimes M$, one may note the  isomorphism of groupoids
\beq G\ltimes (G/H)\cong G\x_H G, \label{isoGR}\eeq
 where the right-hand side is the gauge groupoid 
 of the principal $H$-bundle $G$ with respect to the natural right-action of $H$. 
This isomorphism (given by $(xy\inv, \pi(x))\leftrightarrow [x,y]$) naturally passes to the `algebra of observables,' i.e.\ one has
 \beq
C^*(G\ltimes (G/H))\cong C^*(G\x_H G), \label{isoCG}\eeq
and one may see the right-hand side as a special
case of $C^*(P\x_H P)$ for an arbitrary principal $H$-bundle $P$.\footnote{This generalization is closely related to Kaluza--Klein theory and the  Wong equations; see \cite{Lan}.} Here one has a complete analogue of Mackey's imprimitivity theorem:
the Morita equivalence \beq P\x_H P\sim H \label{MEGL}\eeq at the groupoid level\footnote{The equivalence bibundle is $ P\x_H P\lac P\rac H$, with the given right $H$ action on $P$ and the left action given by $[x,y]y=x$.}
 induces a  Morita equivalence \beq
C^*(P\x_H P)\sim C^*(H) \label{KK} \eeq
 at the \ca ic level, which in turn implies that there is a bijective correspondence between (irreducible) unitary \rep s $U_{\ch}(H)$
and \rep s $\pi^{\ch}(C^*(P\x_H P))$.\footnote{Given  $U_{\ch}(H)$ on a \Hs\ $\H_{\ch}$, the \rep\ $\pi^{\ch}$ is naturally realized on $L^2(P/H,P\x_H \H_{\ch})$, as in the homogeneous case.}

In the old days, the various \irrep s (or superselection sectors) of algebras of observables like $C^*(G\ltimes M)$
or $C^*(P\x_H P)$ were seen as `inequivalent quantizations' of a single underlying classical  system. From this perspective, quantities like spin were seen as degrees of freedom peculiar to and emergent from quantum theory. Starting with geometric quantization in the mid-1960s, however, it became clear that each superselection sectors of said type
is in fact the quantization of a different classical system. The language of Lie groupoids and Lie algebroids allows the most precise and conceptually clearest discussion of this situation. Mathematically, what is at stake here is the relationship between \ncg\ and symplectic geometry as its classical analogue.\footnote{See also \cite{NT} for a different approach to this relationship}
We now turn to this  language. 
\section{Intermezzo: Lie algebroids and Poisson manifolds}\label{LAPM}
Since the notion of a Lie algebroid cannot found in the \ncg\ literature, we provide a complete definition.\footnote{Cf.\ \cite{McK,MM} for  detailed treatments. The concept of a Lie algebroid and the relationship between Lie groupoids and Lie algebroids are originally due to Pradines.} 
\begin{Definition} 
A {\it Lie algebroid} $A$ over a manifold
$M$ is a vector bundle $A \stackrel{\pi}{\raw} M$ equipped with a vector bundle map
$A \stackrel{\al}{\raw} TM$ (called the {\it anchor}), as well as
with a Lie bracket $[\, ,\,]$ on the space  $\cin(M,A)$ of smooth
 sections of $A$, satisfying the Leibniz rule
\beq  [\sg_1,f \sg_2]=f [\sg_1,\sg_2]+ (\al\circ
\sg_1 f) \sg_2\eeq
for all $\sg_1,\sg_2\in\cin(M,A)$ and $f\in\cin(M)$.  
 \end{Definition}
It follows that the map $\sg\mapsto\al\circ\sg: \cin(M,A)\raw \cin(M,TM)$ induced by the anchor is a homomorphism of Lie algebras, where the latter is equipped with the usual commutator of vector fields.\footnote{This homomorphism property used to be part of the definition of a Lie algebroid, but as observed by Marius Crainic it follows from the stated definition.} 

Lie algebroids generalize (finite-dimensional) Lie algebras as well as tangent bundles, and the (infinite-dimensional) Lie algebra $\cin(M,A)$ could be said to be of geometric origin
in the sense that it derives from an underlying finite-dimensional geometrical object. 
Similar to our list of example of Lie groupoids in Section \ref{intermezzo}, one has the following basic classes of \La s.
\begin{itemize}
\item  A {\it Lie algebra}
$\g$, where $A=\g$ and $M$ is a point (which may be identified with the identity element of any Lie group with Lie algebra $g$; see below) and $\al=0$.
\item A {\it manifold} $M$, where $A=M$, seen as the zero-dimensional vector bundle over $M$, evidently with identically vanishing Lie bracket and anchor.
\item The {\it tangent bundle} over a manifold $M$, where $A=TM$ and $\al=\id:TM\raw TM$, with the Lie bracket given by the usual commutator of vector fields. 
\item The {\it gauge algebroid} defined by a principal $H$-bundle $P\raw M$;
here $A=(TP)/H$, so that $\cin(M,A)\cong \cin(M,TP)^H$, which inherits the commutator from
$\cin(M,TP)$ as the Lie bracket defining the algebroid structure, and is equipped with the projection $\al: (TP)/H\raw TM$ induced by  $TP\raw TM$.
\item The {\it action algebroid} $\g \ltimes M$ defined by a $\g$-action on a manifold $M$
(i.e.\ a Lie algebra homomorphism $\g\raw\cin(M,TM)$) has $A=\g\x M$ (as a trivial bundle)
and $\al(X,m)=-\xi_X(m)\in T_mM$. The Lie bracket is 
$$ [X,Y](m)= [X(m),Y(m)]_{\g} +\xi_Y X(m)-\xi_X Y(m). $$
\end{itemize}

It is no accident that these examples exactly correspond to our previous list of \Lg s:
as for groups, any \Lg\ $\Gm$ has an associated \La\ $A(\Gm)$ with the same base space.\footnote{The association $\Gm\raw A(\Gm)$ is functorial in an appropriate way, so that Mackenzie speaks of the {\it Lie functor} \cite{McK}.} Namely, as a vector bundle $A(\Gm)$ is the restriction of $\ker(t_*)$ to $\Gm_0$, and the anchor is $\al=s_*$. One may identify sections of $A(\Gm)$ with left-invariant vector fields on $\Gm$, and under this identification the Lie bracket on $\cin(\Gm_0,A(\Gm))$ is by definition the commutator.

Conversely, one may ask whether a given Lie algebroid $A$ is {\it integrable}, in that it comes from a \Lg\ $\Gm$ in the said way.  That is, is  $A\cong A(\Gm)$ for some \Lg\ $\Gm$?
This is not necessarily the case; see \cite{CF1,Mackenzie95}. 
 
The modern interplay between Lie \Lg s and \La s on the ond hand, and symplectic  geometry on the other is  based on various amazing points of contact. The simplest of these is as follows. 
\begin{Proposition}\label{CouW} {\rm \cite{CDW,Cou}}
The dual vector bundle $A^*$ of a \La\ $A$ is canonically a Poisson manifold.
The Poisson bracket on $\cin(A^*)$ is defined by the following special cases:
$\{f,g\}_{\pm}  = 0$ for $f,g\in\cin(M)$;  $\{\til{\sg},f\}  =
\al\circ\sg f$, where $\til{\sg}\in\cin(A^*)$ is defined by a section $\sg$ of $A$
through the obvious pairing, and finally 
$\{\til{\sg}_1,\til{\sg}_2\}
 =  \widetilde{[\sg_1,\sg_2]}$. 

Conversely, if a vector bundle  $E\raw M$ is a Poisson manifold such that
the Poisson bracket of two linear functions is linear, then $E\cong A^*$ 
for some \La\ $A$ over $M$, with the above Poisson structure.\footnote{This establishes a categorical equivalence between linear Poisson structures
on vector bundles and Lie algebroids. One can also show that in this situation the 
differential forms on $A$ form a differential graded algebra, while those on $A^*\cong E$ (or, equivalently, the so-called polyvector fields on $A$) are a Gerstenhaber algebra; see \cite{Hueb}.}
 \end{Proposition}

The main examples are:
\begin{itemize}
\item  The dual $\g^*$ of a Lie algebra
$\g$ acquires its canonical Lie--Poisson structure (cf.\ \cite{MR}).
\item A manifold $M$, seen as the dual to the zero-dimensional vector bundle $M\raw M$, carries the zero Poisson structure. 
\item A cotangent bundle $T^*M$ acquires the Poisson structure defined by its 
standard symlectic structure. 
\item The dual $(T^*P)/H$ of a gauge algebroid   inherits the canonical Poisson structure
from $T^*P$ under the isomorphism $\cin(T^*P)/H)\cong\cin(T^*P)^H$.
\item The dual $\g^*\ltimes M$ of an
action algebroid acquires the so-called {\it semidirect product}  Poisson structure \cite{KM,MaRaWe}.\footnote{Relative to a basis of $\g$ with structure constants $C_{ab}^c$, this is
 given by $\{f,g\} =C_{ab}^c
\theta_c \frac{\partial f}{\partial\theta_a} \frac{\partial
g}{\partial\theta_b} + \xi_a f \frac{\partial g}{\partial\theta_a} -
\frac{\partial f}{\partial\theta_a}\xi_a g.$ }
\end{itemize}

Combining the associations $\Gm\mapsto A(\Gm)$ and $A\mapsto A^*$, one has an association
\beq \Gm\mapsto A^*(\Gm), \label{class}\eeq
of a Poisson manifold to a \Lg, which we call {\it Weinstein's map}.  As we shall see, this  is a classical analogue of Connes's map \er{Connesmap} in every possible respect.
\section{Symplectic groupoids and the category of \PM s}
Another important point of contact between Poisson manifolds and \La s that is relevant for what follows is the following construction.
\begin{Proposition}\label{PTP}{\rm \cite{CDW}}
If $P$ is a Poisson manifold, then $T^*P$ is canonically a \La\ over $P$.
\end{Proposition}
The anchor is just the usual map $T^*P\raw TP$, $\al\mapsto\al^{\sharp}$ (e.g., $df\mapsto X_f$)\footnote{The  Hamiltonian vector field $X_f$ defined by a smooth function $f$ on a Poisson manifold $P$ is defined by $X_f g=\{f,g\}$.} defined by the Poisson structure, whereas the Lie bracket is
\beq [\al,\bt]=\mathcal{L}_{\al^{\sharp}}\bt -\mathcal{L}_{\bt^{\sharp}}\al + d\pi(\al,\bt),
\eeq
where $\pi$ is the Poisson tensor. Combining this with Proposition \ref{CouW}, one infers that $TP$ is a Poisson manifold whenever $P$ is.\footnote{In addition, one may recover the Poisson cohomology of $P$ as the \La\ cohomology of $T^*P$ \cite{McK,WeinsteinXu}.}

The following definition will play a key role for us in many ways.
\begin{Definition}\label{Defint}{\rm \cite{CDW}}
A Poisson manifold $P$ is called {\rm integrable} when the associated \La\ $T^*P$
 is integrable (in being the \La\ of some \Lg).
\end{Definition}

If $P$ is an integrable \PM, a groupoid $\Gm(P)$ for which $A(\Gm(P))\cong T^*P$ 
(and hence $\Gm(P)_0\cong P$) turns out to have the structure of a {\it symplectic groupoid}.
\begin{Definition}\label{DEfSG} {\rm \cite{Kar,Wei87,Zak}}
A {\rm symplectic groupoid} is  a \Lg\ whose total space $\Gm_1$ is a symplectic manifold, such that the 
 graph of $\Gm_2\subset \Gm\x \Gm$ is a Lagrangian submanifold of $\Gm\x \Gm\x \Gm^-$.
\end{Definition}
See also \cite{CDW,McK,MiW}. Symplectic groupoids have many amazing properties, and in our opinion their introduction into symplectic geometry has been  the biggest leap forward since the subject was founded.\footnote{It would be tempting to say that a suitable analogue of a symplectic groupoid has not been found in \ncg\ so far, but in fact an analysis of the categorical significance of 
symplectic groupoids, \PM s, and operator algebras \cite{Lan3} shows that the `quantum symplectic groupoid' associated to a \ca\ $A$ is just $A$ itself, whereas for a von Neumann algebra its standard form plays this role.} For example:
\begin{enumerate}
\item There exists a unique Poisson structure on $\Gm_0$ such that
$t$ is a Poisson map and $s$ is an anti-Poisson map.
\item $\Gm_0$ is a Lagrangian submanifold of $\Gm_1$.
\item The inversion in $\Gm$ is an anti-Poisson map.
\item The foliations of $\Gm$ defined by the
levels of $s$ and $t$ are mutually symplectically orthogonal.
\item If $\Gm$ is s-connected,\footnote{This means that each fiber $s\inv(u)$ is  connected, $u\in \Gm_0$. Similarly for s-simply connected.} then $s^*\cin(\Gm_0)$ and
$t^*\cin(\Gm_0)$ are each other's Poisson commutant.
\item The symplectic leaves of $\Gm_0$ are the connected components of the 
$\Gm_1$-orbits.
\end{enumerate}
With regard to the first point, the Poisson structure on $\Gm(P)_0$ induces the given one on $P$ under the diffeomorphism $\Gm(P)_0\cong P$. For later use, we record:
\begin{Proposition}\label{SGSC} {\rm \cite{CDW,CF2,Lan3}}
If a \PM\ $P$ is integrable, then there exists an s-connected and s-simply connected symplectic groupoid $\Gm(P)$
over  $P$, which  is unique up to isomorphism.
\end{Proposition}

For example, suppose that $\Delta$ is a \Lg;
is the Poisson manifold $A^*(\Delta)$ it defines by \er{class} integrable? The answer is yes, and one may take
\beq \Gm(A^*(\Delta))=T^*\Delta, \eeq
the so-called {\it cotangent groupoid} of $\Dl$ \cite{CDW} (see also \cite{Lan4,McK}). 
This is s-connected and s-simply connected iff $\Dl$ is.

Using the above constructions, we now define a category $\GP$ of \PM s, which will play a central role in what follows. 
First, the objects of $\mathfrak{P}$ are
{\it integrable} Poisson manifolds; the integrability condition turns out to be necessary in order to have identities in $\GP$; see below. 
In the spirit of general Morita theory \cite{Faith73}, the arrows in $\GP$ are bimodules in an appropriate sense. Bimodules for \PM s are known as {\it dual pairs} \cite{Kar89,Wei83}. A  dual pair $Q\law S \raw P$ consists of a symplectic manifold $S$, Poisson manifolds $Q$ and $P$, and complete Poisson maps $q:S\raw Q$ and $p:S\raw P^-$, such that $\{q^* f, p^* g\}=0$ for all $f\in\cin(Q)$ and $g\in\cin(P)$. To explain the precise class of dual pairs whose isomorphism classes form the arrows in $\GP$, we need a symplectic analogue $\GS$ of the category $\GG$ (cf.\ Definition \ref{CatG}). In preparation, we call an action of a symplectic groupoid $\Gm$ on a symplectic manifold $S$ {\it symplectic}  when the graph of the action in $\Gm\x S\x S^-$ is Lagrangian \cite{CDW,MiW}. 
\begin{Definition}{\rm \cite{Lan3}}
The category $\GS$ is the subcategory of the category $\GG$ (of Lie groupoids and principal bibundles) whose objects are symplectic groupoids and whose arrows are isomorphism classes of principal bibundles for which
the two groupoid actions are symplectic.
\end{Definition}
We call such bibundles {\it symplectic}.
As we have seen (cf.\ Section \ref{LAPM}), the base space of a symplectic groupoid is a \PM. Moreover, it can be shown \cite{CDW,MiW} that the base map $S\raw \Gm_0$ of a symplectic action of a symplectic groupoid $\Gm$ on a symplectic manifold $S$  is
a complete Poisson map such that for $(\gm,y)\in \Gm\times^{s,\rh}_{\Gm_0}S$ 
with $\gm=\phv_1^{t^*f}(\rh(y))$,
one has $\gm y=\phv^{\rh^*f}_1(y)$ (here $\phv_t^g$ is the Hamiltonian
flow induced by a function $g$, and $f\in\cin(\Gm_0)$).
Conversely, when $\Gm$ is s-connected and
s-simply connected, a given complete Poisson map $\rh:S\raw \Gm_0$
is the base map of a unique symplectic $\Gm$ action on
$S$ with the above property \cite{Xu1}. Furthermore, it is easy to show that
the base maps of a symplectic bibundle form a dual pair.
We call a dual pair arising from a symplectic principal bibundle in this way {\it regular}.
\begin{Definition}\label{CatP}
The objects of the category $\GP$ of Poisson manifolds and dual pairs are integrable \PM s, and its arrows are isomorphism classes of regular dual pairs.
\end{Definition}
The identities in $\GP$ are $1_P=[P\law \Gm(P)\raw P]$, where $\Gm(P)$ is ``the'' s-connected and s-simply connected symplectic groupoid over $P$; cf.\ Proposition \ref{SGSC}. As in every decent version of Morita theory, isomorphism of objects in $\GP$ comes down to Morita equivalence of \PM s (in the sense of Xu \cite{Xu1}).

It is clear  that $\GP$ is equivalent to the full
subcategory  $\GS_c$ of $\GS$ whose objects are s-connected and s-simply connected
symplectic groupoids; the advantage of working with $\GP$ rather than $\GS_c$ lies both in the greater intuitive appeal of Poisson manifolds and dual pairs over symplectic groupoids and symplectic principal bibundles, and also in the fact that the composition of arrows can be 
formulated in direct terms (i.e.\ avoiding arrow composition in $\GS$ or $\GG$) using a generalization of the familiar procedure of symplectic reduction \cite{Lan3,Xu2}.

 For example, a strongly Hamiltonian group action $G\lac S$ famously defines a dual pair $$S/G \stackrel{\pi}{\law}S\stackrel{J}{\raw}\mathfrak{g}^*$$ (where $J$ is the momentum map of the action) \cite{Wei83}, whose
product with the dual pair $\g^* \hookleftarrow 0\raw pt$ in $\GP$ equals 
$S/G \hookleftarrow S/\hspace{-1mm}/G\raw pt$ (if we assume $G$ connected). In other words, the Marsden--Weinstein quotient $S/\hspace{-1mm}/G$ \cite{AM,MR} may be interpreted in terms of the category $\GP$ (see Section \ref{GSC} below for the significance of this observation.)
\section{The classical imprimitivity theorem}
There is a complete classical analogue of Mackey's theory of imprimitivity for (Lie) group actions \cite{GSbook,Lan,Ziegler}. Firstly, the classical counterpart of a \rep\ of a \ca\ on a \Hs\ is a so-called {\it realization} of a Poisson manifold  $P$ on a symplectic manifold $S$ \cite{Wei83}; this is a {\it complete} Poisson map $S\stackrel{\rh}{\raw} P$.\footnote{Some authors speak of a realization in case that $\rh$ is surjective, but not necessarily complete. The completeness of $\rh$ means that the Hamiltonian vector field $X_{\rh^*f}$ on $S$ has a complete flow
for each $f\in\cci(P)$ (i.e.\ the flow is defined for all times). This condition turns out to be the classical counterpart of the requirement that $\pi(a)^*=\pi(a^*)$ for \rep s of a \ca.
The analogy between completeness of the flow of a vector field and self-adjointness of an operator is even more powerful in the setting of unbounded operators; for example,
the Laplacian on a Riemannian manifold $M$ is essentially self-adjoint on $\cci(M)$
  when $M$ is geodesically complete \cite{AM}.} The appropriate symplectic notion of irreducibility is that $$\{X_{\rh^*f}(x)\mid f\in\cin(P)\}=T_xS$$ for all $x\in S$ (where $X_g$ is the Hamiltonian vector field of $g\in\cin(S)$); it is easy to show (cf.\ Thm.\ I.2.6.7 in \cite{Lan}) that $\rh$ is irreducible iff $S$ is symplectomorphic to a covering space of a symplectic leaf of $P$ (and $\rh$  is the associated projection followed by injection).
In particular, any Poisson manifold has at least one  irreducible realization.\footnote{The appropriate symplectic notion of faithfulness is simply that $\rh$ be surjective; it was recently shown by Crainic and Fernandes \cite{CF2} that a Poisson manifold admits a faithful realization iff it is  integrable; cf.\ Definition \ref{Defint}. Along with their solution of this integrability problem \cite{CF1}, this is one of the deepest results in symplectic geometry to date.} 

Secondly, we provide the classical counterpart of Definition \ref{soi}. It goes without saying that in the present context $G$ is a Lie group and $M$ a manifold, all actions being smooth by definition. 
\begin{Definition}\label{csoi}
Given a $G$-action on $M$, a {\rm $G$-covariant realization of $M$} (seen as a Poisson manifold with zero Poisson bracket) is a complete Poisson map $S\stackrel{\rh}{\raw} M$, where $S$ is a symplectic manifold equipped with a strongly Hamiltonian $G$-action,\footnote{In the sense that the $G$-action has an equivariant momentum map $J:S\raw\g^*$ \cite{AM,MR}.}
and $L_x(\rh^*f)=\rh^* L_x(f)$ for all $f\in\cin(M)$.
\end{Definition}
The significance of this definition and its analogy to Definition \ref{soi} are quite obvious; instead of a \rep\ $\pi:C_0(M)\raw B(\H)$ one now has a Lie algebra homomorphism
$\rh^*:\cin(M)\raw\cin(S)$. 
Its relationship to the material in the preceding section is as follows:
\begin{Proposition}{\rm \cite{Xu2}}
When $G$ is connected, a $G$-covariant realization of $M$ may equivalently be defined as
a realization $S\stackrel{\sg}{\raw} \g^*\ltimes M$ (equipped with the semidirect product Poisson structure) whose associated $\g$-action on $S$ is integrable (i.e.\ to a $G$-action on $S$).
\end{Proposition}
The $\g$-action on $S$ in question is given by $X\mapsto X_{\sg^*\til{X}}$, where
$X\in\g$ defines a linear function $\til{X}:\g^*\raw\C$ by evaluation (and consequently also defines a function on $\g^*\x M$ that is constant on $M$, which we denote by the same symbol). Of course, given $S\stackrel{\rh}{\raw} M$ as in Definition \ref{csoi}, one defines $S\stackrel{\sg}{\raw} \g^*\ltimes M$ by $\sg=(J,\rh)$; the nontrivial part of the proposition lies in the completeness of $\sg$, given the completeness of $\rh$. 

One then has the following classical analogue of Mackey's imprimitivity theorem. \begin{Theorem}\label{climthm} {\rm \cite{Ziegler}}
 Up to symplectomorphism, there is a bijective correspondence between $G$-covariant realizations $S\stackrel{\rh}{\raw} G/H$ of $G/H$ (with zero Poisson structure) and strongly Hamiltonian $H$-spaces
$S_{\rh}$, as follows:
\begin{itemize}\item
Given $S_{\rh}$, the Marsden--Weinstein quotient (at zero) $S^{\rh}=(T^*G\x S_{\rh})/\hspace{-1mm}/H$ 
is a $G$-covariant realization of $G/H$.\footnote{The $G$-action  inherited from the $G$-action on $T^*G$ is given by pullback of left-multiplication, and the map $S^{\rh}\raw G/H$ is  inherited from the natural map $T^*G\raw G\raw G/H$.}
\item Conversely, given $S\stackrel{\rh}{\raw} G/H$ there exists a strongly Hamiltonian $H$-space $S_{\rh}$ such that $S\cong S^{\rh}$.
\end{itemize}
This correspondence preserves irreducibility.

When $G$ is connected, this correspondence may be seen as being between realizations $S\stackrel{\sg}{\raw} \g^*\ltimes (G/H)$  whose associated $\g$-action on $S$ is integrable, and realizations $S_{\rh}  \stackrel{J_{\rh}}{\raw}\h^*$ whose
associated $\h$-action on $S_{\rh}$ is integrable. 
\end{Theorem}

The original proof of this theorem was lengthy and difficult \cite{Lan,Ziegler}. Fortunately,   as in the quantum case, there exists a direct categorical argument, according to which at least the last part of Theorem \ref{climthm} is a consequence of \er{GHGH} as well. Namely, the following analogue of Theorem \ref{Lan2Thm} holds:
\begin{Theorem} \label{Lan2bis} {\rm \cite{Lan2}}
Weinstein's map $\Gm\mapsto A^*(\Gm)$ is functorial from $\mathfrak{G}_c$ to $\mathfrak{P}$.
\end{Theorem}
Recall that $\mathfrak{G}_c$ is  the full subcategory of $\mathfrak{G}$ whose objects are s-connected and s-simply connected Lie groupoids, and that the category $\GP$
of Poisson manifolds and dual pairs  has been  defined in the previous section. For example, $G\x (G/H)$ is an object in  $\mathfrak{G}_c$ iff $G$ is connected and simply connected. Assume this to be the case for the moment. As already mentioned, the category $\mathfrak{P}$ has a feature analogous to the category $\mathfrak{C}$ of \ca s, namely that two objects are isomorphic iff they are Morita equivalent Poisson manifolds in the sense of Xu \cite{Xu1}. Consequently, 
similar to Corollary \ref{Lan4cor} one has:
\begin{Corollary}\label{PME}  {\rm \cite{Lan4}}
Weinstein's  map $\Gm\mapsto A^*(\Gm)$ preserves Morita equivalence, 
in the sense that if $\Gm$ and $\Dl$ are Morita equivalent  s-connected and s-simply connected Lie groupoids, then $A^*(\Gm)$ and $A^*(\Dl)$ are Morita equivalent 
 Poisson manifolds in the sense of Xu.
\end{Corollary}
Thus
the Morita equivalence \er{GHGH} of Lie groupoids implies the Morita equivalence 
\beq \g^*\ltimes (G/H)\sim \h^* \label{cGHGH}\eeq
of Poisson manifolds. As for \ca s (and algebras in general), if two Poisson manifolds $P_1, P_2$ are Morita equivalent, then they have equivalent categories of realizations, and the equivalence bimodule implementing this Morita equivalence comes with an explicit procedure that defines a realization of $P_2$ given one of $P_1$, and vice versa. This procedure is a certain generalization of symplectic reduction \cite{GSbook,Lan,Xu1} (much as the corresponding Rieffel induction procedure for \ca s is a generalization of Mackey induction). In the case at hand, viz.\ \er{cGHGH}, this precisely gives the prescription 
stated in Theorem \ref{climthm},  proving its  last part at least for simply connected $G$. If $G$ fails to be simply connected, one passes to its universal cover $\til{G}$, and lets it act on $G/H$ via the projection $\til{G}\raw G$. Hence $G/H\cong  \til{G}/\hat{H}$ for some
$\hat{H}\subset  \til{G}$; Lie theory gives $\til{\g}=\g$ and $\hat{\h}=\h$. The conclusion
\er{cGHGH} still follows, this time as a consequence of $\til{G}\ltimes (\til{G}/\hat{H})\sim \hat{H}$ rather than of \er{GHGH}.

We state a rather satisfying classical analogue of Proposition \ref{Glimm}, which is essentially a corollary to Theorem \ref{climthm}.
\begin{Proposition}\label{slactionpa} {\rm \cite{MaRaWe}}
The symplectic leaves of of the semidirect Poisson structure on $\g^*\ltimes M$ are classified by pairs $(\CO,\CO')$, where $\CO$ is a $G$-orbit in $M$, 
and $\CO'$ is a coadjoint orbit of the stabilizer of an arbitrary point in $\CO$. \end{Proposition}
If we call the stabilizer in question $H$, 
 the symplectic leaf $L_{(\CO,\CO')}$ corresponding to the pair $(\CO,\CO')$ is given by \beq
L_{(\CO,\CO')}=\{(\theta,q)\in \g^*\x Q\,|\, q\in\CO,\,
(-\Co(s(q)\inv)\theta \rst \h^*)\in \CO'\}, \eeq where $s:\CO\simeq
G/H\raw G$ is an arbitrary section of the canonical principal $H$-bundle $G$ over $G/H$, and $\Co$ is the coadjoint action of $G$ on $\g^*$.

Furthermore, one has a classical counterpart of \er{isoCG}, namely 
 an isomorphism
\beq \g^*\ltimes (G/H) \cong (T^*G)/H \eeq
of Poisson manifolds. This may be generalized from the principal $H$-bundle $G$ to arbitrary principal $H$-bundles $P$, provided that $P$ is connected and simply connected
(this assumption was not necessary in the quantum case). In that case,  we may apply Corollary \ref{PME}  to find  a Morita equivalence of Poisson manifolds
\beq  (T^*P)/H \sim \h^*. \eeq
\section{Deformation quantization}\label{DQsection}
Largely due to the functoriality of Connes's map \er{Connesmap} and its classical counterpart \er{class}, we have observed a striking analogy between the \ca\ $C^*(\Gm)$ and the \PM\ $A^*(\Gm)$ associated to a \Lg\ $\Gm$. Beyond an analogy, the classical object $A^*(\Gm)$ turns out to be related to its quantum counterpart through deformation quantization in the \ca ic setting proposed by Rieffel:
\begin{Definition}\label{gsq}{\rm \cite{Rie89,Rie94}}
A {\rm \ca ic deformation quantization} of a Poisson manifold $P$ is a
continuous field of \ca s $(A,A_{\hbar})_{\hbar\in [0,1]}$,\footnote{Here $A$ is the \ca\ of  sections of the given field, which defines its continuity structure. A continuous field $(A,A_x)_{x\in X}$ of \ca s comes with surjective morphisms $\pi_x:A\raw A_x$.
} where
$A_0=C_0(P)$, with a Poisson algebra $\til{A}_0$ densely
contained in $C_0(P)$ and a cross-section $Q:\til{A}_0\raw A$ of
$\pi_0$, such that, in terms of $Q_{\hbar}=\pi_{\hbar}\circ Q$, 
for all $f,g\in \til{A}_0$ one has
\begin{equation}
\lim_{\hbar\rightarrow 0} 
\|\frac{i}{\hbar}[Q_{\hbar}(f),Q_{\hbar}(g)]-Q_{\hbar}(\{f,g\})\|_{\hbar} =0. \label{Dirac}
\end{equation}
\end{Definition}
This has turned out to be an fruitful definition of quantization (cf.\ \cite{Lan}). In many interesting examples the fiber algebras are non-isomorphic even away from $\hbar=0$
(cf.\ \cite{Rie89,Rie94} and Footnote \ref{CLfn} below), but  
in the case at hand the situation is simpler.\footnote{Technically, the field in Theorem
\ref{LRThm} is  said to be trivial away from
$\hbar=0$, in the sense that $A_{\hbar}=B$ for all $\hbar\in (0,1]$
and one has a  short exact sequence $0\raw CB\raw A\raw A_0\raw 0$ (where
 $CB=C_0((0,1],B)$ is the cone of $B$).} 
\begin{Theorem}\label{LRThm} {\rm \cite{Lan1,LR,Ram}}\footnote{See also \cite{NWX} for a version of this result in the setting of formal deformation quantization (i.e.\ star products), and also cf.\ \cite{Raca}.
} For any \Lg\ $\Gm$, the field
 $A_0=C_0(A^*(\Gm))$, $A_{\hbar}=C^*(\Gm)$ for $\hbar\neq 0$, and $A=C^*(\Gm^T)$,
the \ca\ of the tangent groupoid $\Gm^T$ of $\Gm$,\footnote{Following Connes's definition of the special case of the pair groupoid $\Gm=M\x M$ around 1980 (see \cite{Con}), the tangent groupoid (or adiabatic groupoid) of an arbitrary Lie groupoid was independently defined in \cite{HS,Wei89}. See also \cite{Lan,Pat}.}
 defines a \ca ic deformation quantization of $A^*(\Gm)$.\footnote{The same statement holds for the corresponding reduced groupoid \ca s.}
\end{Theorem}

We refer to the literature cited for the specification of $\til{A}_0$, as well as for the proof of \er{Dirac}. The proof of the remainder of the theorem actually covers a much more general situation, as follows \cite{Ram}.\footnote{This setting was originally suggested by Skandalis.}
\begin{Definition}\label{field}
A {\rm field of Lie groupoids} is a triple $(\SG,X,p)$, with $\SG$ a Lie groupoid,
 $X$ a manifold, and $p:\SG\rightarrow X$ a surjective submersion such that $p=p_0\circ r=p_0
\circ s$, where $p_0=p\rst\SG_0$. \end{Definition}

It follows that each $\SG_x=p\inv(x)$ is a Lie subgroupoid of $\SG$
over  $\SG_0\cap p\inv(x)$, so that
$\SG=\coprod_{x\in X} \SG_x$ as a groupoid.  One may then
form the convolution \ca s $C^*(\SG)$ and $C^*(\SG_x)$. Each $a\in
C_c(\SG)$ (or $\cci(\SG)$) defines $a_x=a\rst \SG_x$ as an element of
$C_c(\SG_x)$ (etc.). These maps $C_c(\SG)\raw C_c(\SG_x)$ are continuous in
the appropriate norms, and extend to maps $\pi_x: C^*(\SG)\raw
C^*(\SG_x)$.  Hence one obtains a field of \ca s
\beq (A=C^*(\SG),A_x=C^*(\SG_x))_{x\in X} \label{field2}\eeq over $X$, where $a\in C^*(\SG)$
defines the section $x\mapsto \pi_x(a)$.\footnote{A similar statement applies
to the corresponding reduced \ca s.}
 The question now arises when this field is continuous.
\begin{Lemma}\label{ramlem}{\rm \cite{Ram}}
The field  \er{field2} is continuous at all points
where $\SG_x$ is amenable \cite{AR,Ren}.\footnote{And similarly for the case of
reduced \ca s.}
\end{Lemma}
For example, the tangent groupoid $\Gm^T$ of a given \Lg\ $\Gm$ forms a field of \Lg s over $[0,1]$,
with $\Gm^T_0=A(\Gm)$ (seen as a Lie {\it groupoid} instead of a \La\ in the way every vector bundle $E\stackrel{\pi}{\raw} M$ defines a \Lg\ over its base space, namely by $s=t=\pi$ and fiberwise addition) and $\Gm^T_{\hbar}=\Gm$ for $\hbar\in (0,1]$. 
This eventually implies  Theorem \ref{LRThm} (except for \er{Dirac}); the same strategy also leads to far-reaching generalizations thereof.\footnote{Lemma \ref{ramlem} applies much more generally to fields of locally compact groupoids. In the context of \ca ic deformation quantization, there are two typical situations. In the smooth (Lie) case studied in this paper, all $\SG_{\hbar}$ are the same for $\hbar\neq 0$ but possibly not amenable, whereas $\SG_0$ is amenable. The former property
then yields continuity at $\hbar=0$ by the lemma, whereas the latter
gives continuity on $(0,1]$. In the context of Definition \ref{gsq},
the reason why $G_0$ is amenable is that  $A_0$ must be commutative,
which implies that $G_0$ is a bundle of abelian groups. But such
groupoids are always amenable \cite{AR}. In the \'{e}tale case
 all $\SG_{\hbar}$ are typically different from each other, but they are all amenable.
See \cite{Cad} for a description of  noncommutative
tori  and the noncommutative four-spheres of Connes and
Landi \cite{ConnesLandi} (and of many other examples) as deformation quantizations
along these lines.\label{CLfn}} 

In physics, Theorem \ref{LRThm} describes the quantization of particles with both internal and spatial degrees of freedom in a very wide setting. 
In noncommutative geometry, certain constructions of Connes in index theory turn out to be special cases of Theorem \ref{LRThm}.\footnote{One instance is the map $p!: K^*(F^*)\raw K_*(C^*(V,F))$ on p.\ 127 of \cite{Con}, which plays a key role in the definition of the analytic assembly map for foliated manifolds. This is the K-theory map
 induced by the continuous field of Theorem \ref{LRThm}, where $\Gm$ is
the holonomy groupoid of the foliation.
 The index groupoid for a vector bundle map $L:E\raw F$ defined in
 \cite[\S II.6]{Con} is another example. Here one has a Lie groupoid $\Gm=\mathrm{Ind}_L=F\rtimes_L
 E$ over $F$, whose Lie algebroid is $F\x_B E$. This is a vector
 bundle over $B$, and in the above formalism it should be regarded as
 a groupoid over $F$ under addition in each fiber. Hence
 $A_0=C^*(F\x_B E)\cong C_0(F\x E^*)$. The corresponding K-theory map
occurs in Connes's construction of the Gysin map $f_{!}:
 K^*(X)\raw K^*(Y)$ induced by a smooth map $f:X\raw Y$ between
 manifolds.} As to the ideology of \ncg, the theorem shows that the two fundamental classes of noncommutative manifolds, namely the ones defined by a singular quotient and the ones defined by deformation \cite{Con,Con2000}, overlap. For in case that the equivalence relation defining the quotient in question can be codified by a Lie groupoid $\Gm$, the noncommutative space $C^*(\Gm)$ associated with the quotient space is at the same time a deformation of the dual of its \La.

Furthermore, Connes's philosophy in dealing with singular quotients, and especially his description of the Baum--Connes conjecture in Ch.\ II of \cite{Con}, actually suggests a procedure for the quantization of such spaces. We explain this in a simple example \cite{Lan5}. Suppose a Lie group $G$ acts on a manifold $M$; it acts on $T^*M$ by pull-back, and we happen to be interested in quantizing the quotient $(T^*M)/G$. In case that the $G$-action is free and proper the situation is completely understood: the quotient is a \PM\ of the type $A^*(\Gm)$ for $\Gm=M\x_G M$, 
to which Theorem \ref{LRThm} applies (see also \cite{Lan} for a detailed study of this case).
However, if the $G$-action is not free (but still assumed to be proper), the quotient $(T^*M)/G$ may fail to be a manifold, let alone a \PM. According to Connes, one should replace the space $(T^*M)/G$ by the 
groupoid $T^*M\rtimes G$, and regard the associated noncommutative space $C^*(T^*M\rtimes G)$ as a {\it classical} space. If the $G$-action {\it is} free, one has a Morita equivalence of Lie groupoids
\beq T^*M\rtimes G\sim (T^*M)/G \label{TMME}\eeq
which by Corollary \ref{Lan4cor} implies a Morita equivalence
\beq C^*(T^*M\rtimes G)\sim C^*((T^*M)/G) \label{RME}\eeq
of \ca s.\footnote{See \cite{Rie82} for the original, non-groupoid proof  of \er{RME}.} 
In general, we propose to quantize  the singular space $(T^*M)/G$ by deforming $C^*(T^*M\rtimes G)$, which may be done by the field of \Lg s defined by the tangent groupoid $\Gm^T$ of $\Gm=(M\x M)\rtimes G$. This field has fibers $\Gm^T_0=TM\rtimes G$
(where $TM$ is seen as a \Lg, as explained above), and $\Gm^T_{\hbar}=(M\x M)\rtimes G$. By Lemma \ref{ramlem} (which applies because $TM\rtimes G$ is amenable; see Lemma 2 in \cite{Lan5}), this field of groupoids leads to a continuous field of \ca s with $A=C^*(\Gm^T)$, etc., in the   familiar way. The fibers of the latter field are simply $A_0  =  C_0(T^*M)\rtimes G$ and 
$A_{\hbar}  =  K(L^2(M))\rtimes G$ for all $\hbar\in(0,1]$. To what extent this reflects physical desiderata remains to be seen.
\section{Functorial quantization}\label{FQ}
The final application of groupoids to physics and \ncg\ we wish to describe in this paper is a functorial approach to quantization. In our opinion this forms the natural outcome of the categorical approach to Mackey's imprimitivity theorem described above. 
Beyond the desire to complete Mackey's program, why should one wish to turn quantization into a functor? Historically, \qm\ started with Heisenberg's paper
{\it \"{U}ber die quantentheoretische Umdeutung kinematischer und mechanischer Beziehungen}\footnote{{\it On the quantum-theoretical reinterpretation of kinematical and mechanical relations}.}\cite{Heis}.
One might argue that the proper mathematical reading of Heisenberg's idea of {\it Umdeutung} ({\it reinterpretation}) is that the transition from classical to \qm\ should be given by a functor. Indeed, attempts to make quantization functorial date back at least to van Hove's famous paper from 1951 \cite{vanH}
(see also \cite{Got1,Got2}), the general conclusion being that functorial quantization is impossible (see \cite{Lan6} and refs.\ therein). However,
all no-go theorems in this direction start from wrong and naive categories, both on the classical and on the quantum side.  

Instead, though we have to warn the reader that we are presenting a program rather than a theorem here, it seems possible to interpret quantization as a functor $\CQ$ from either the category $\GS$  (cf.\  Definition \ref{CatG}), or, more straightforwardly, from the category $\GP$ (see Definition \ref{CatP}; recall that $\GP$ is equivalent to a full subcategory of $\GS$) to the category $\KK$ defined by Kasparov's bivariant K-theory (see \cite{Bla,Con}).\footnote{The objects of $\KK$ are separable \ca s, and the arrows are $\mathrm{Hom}_{\KK}(A,B)=KK(A,B)$, composed with Kasparov's product $KK(A,B)\x KK(B,C)\raw KK(A,C)$.} 
 This was first proposed in \cite{Lan6,Lan7,Lan8}.
 Beyond the defining property of making quantization functorial, this program would:
\begin{itemize}
\item  Unify deformation quantization and
geometric quantization into a single operation (the former becoming the object side of the quantization functor and the latter the arrow side);
 \item Imply the functoriality of shriek maps in K-theory \cite{AS1},
in particular providing a natural home for Connes-style proofs and generalizations of index theorems \cite{Con,CS};
 \item Imply the ``quantization commutes with reduction''  conjecture of Guillemin and Sternberg \cite{GS};
 \item Provide unlimited generalizations of this conjecture, e.g., to
 noncompact Lie  groups and Lie groupoids (see \cite{HL} for the former).
 \end{itemize}
 
It should be clear that the use of groupoids is essential in this program, since the classical category $\GS$ of symplectic groupoids and principal  symplectic bibundles either forms the domain of the quantization functor $\CQ$, or, in case one more naturally starts from $\GP$,  plays an essential  role in the definition of the latter category. 

Let us indeed construe quantization as a functor $\CQ:\GP\raw KK$.  This means that quantization sends (isomorphism  classes of)
dual pairs into (homotopy  classes of)  Kasparov  bimodules. More precisely, 
 if Poisson manifolds $P_1$ and $P_2$ are quantized by
(separable) \ca s $\CQ(P_1)$ and $\CQ(P_2)$, respectively, a dual pair $P_1\law M\raw P_2$ should be quantized by an element 
\beq \CQ(P_1\law M\raw P_2)\in KK(\CQ(P_1),\CQ(P_2)), \label{QPM}\eeq
where $KK(-,-)$ is the usual Kasparov group \cite{Bla,Con}.
Roughly speaking, the construction of $\CQ(P)$ should be done by some \ca ic version of deformation quantization, whereas that of $\CQ(P_1\law M\raw P_2)$
should come from a far-reaching generalization of geometric quantization first proposed, in special cases, by Raoul Bott;\footnote{This was done in seminars and conversations; no paper by Bott containing his proposal seems to exist. (V. Guillemin and R. Sjamaar, private communications.)}  see \cite{GGK,Sja}.
This proposal turns out to be closely related to Connes's construction of shriek maps \cite{Con,CS}.

To explain the construction of \er{QPM}, we assume that the symplectic manifold $(M,\om)$ is prequantizable. Cf.\ \cite{GGK,Par} for details of the following approach to geometric quantization. One picks an almost complex structure $\CJ$ on $M$ that is compatible with $\om$ (in that $\om(-,\CJ-)$ is positive definite and symmetric). This $\CJ$ canonically induces 
a $\spinc$ structure on $TM$, which should subsequently be twisted by 
a prequantization line bundle $L$ line bundle over $M$ to obtain a $\spinc$ structure  $(P,\cong)$ on $M$.\footnote{We here define a  $\spinc$ structure  on $M$ as an equivalence class of principal $\spinc(n)$-bundle $P$ over $M$  with an isomorphism $P\x_{\pi} \R^n\cong TM$ of vector bundles. Here  $n=\dim(M)$ and the bundle on the left-hand side is the bundle associated to $P$ by the defining \rep\ of $SO(n)$. Connes's construction of shriek maps lacks the twisting with the prequantization line bundle.}  Denote the  (complex)
spin \rep\ of $\spinc(n)$ on the finite-dimensional Hilbert space $S_n$
by $\Delta_n$. One may then form the  associated spinor bundle
$\CS_n=P\x_{\Dl_n}S_n$, with Dirac operator $\DS:\cin(M,\CS_n)\raw \cin(M,\CS_n)$. For even $n$ (the case that applies here, as $M$ is symplectic) the spin \rep\ decomposes into two
irreducibles $\Dl_n=\Dl_n^+ \oplus \Dl_n^-$ on $S_n=S_n^+\oplus S_n^-$, so that also the vector bundle $\CS_n$ decomposes accordingly as $\CS_n=\CS_n^+\oplus \CS_n^-$. 
Being odd with respect to this decomposition, the Dirac operator then splits accordingly 
as $\DS^{\pm}=\cin(M,\CS^{\pm})\raw \cin(M,\CS^{\mp})$.

Given a dual pair $P_1\law M\raw P_2$, the fundamental idea is to use the map $M\raw P_2$ to turn the appropriate completion of $\cci(M,\CS_n)$ to a graded Hilbert $C^*(\CQ(P_2))$ module $\CE$, and subsequently, to use the map $P_1\law M$ to construct an action of $C^*(\CQ(P_1))$ on $\CE$, producing a $C^*(\CQ(P_1))$-$C^*(\CQ(P_2))$ graded Hilbert bimodule. The final step is to employ  the Dirac operator $\DS$ to enrich this bimodule into a Kasparov cycle, whose homotopy class  defines the element \er{QPM} we are after. 

This procedure has so far been carried through in a few cases only, namely those in which Theorem \ref{LRThm} states how the \PM s $P_i$ are to be quantized, and in which simultaneously techniques from the literature on the Baum--Connes conjecture \cite{BCH,Con,Val2} are available to construct   \er{QPM} according to the procedure just sketched. 
The simplest case is $P_1=P_2=pt$ (i.e.\ a point) and $M$ an arbitrary compact prequantizable symplectic manifold.\footnote{Let us note that the associated dual pair $pt\law M\raw pt$ does not define an element of our category $\GP$, but this nuisance does not stop us from proceeding.} Most people would agree that $\CQ(pt)=\C$, and under the isomorphism $KK(\C,\C)\cong\Z$ the Kasparov cycle defined by $\DS$ is just the Fredholm index of $\DS_+$ \cite{Bla}. This number, then, is Bott's quantization of $(M,\om)$. Consequently, we have
\beq \CQ(pt\law M\raw pt)=\mathrm{Index}(\DS_+).\label{Bott1}\eeq
 \section{Quantization commutes with reduction}\label{GSC}
The above
 definition of quantization gains in substance when one passes to a dual pair $M/G \law M\raw \g^*$ defined by a strongly Hamiltonian group action $G\lac M$ in the usual way \cite{Wei83}. For simplicity, we will actually use the dual pair $pt\law M\raw \g^*$.\footnote{This dual pair does not define an element of $\GP$, but  this does not affect any of our arguments.} Theorem \ref{LRThm} tells us that $\CQ(\g^*)=C^*(G)$, where $G$ is any Lie group with Lie algebra $G$; we take the connected and simply connected one.\footnote{Here the use of the category $\GS$ as the domain of the quantization functor $\CQ$ is more satisfactory. The classical data is then formed by the $G$-action on $M$ itself (in the guise of the associated symplectic action of the symplectic groupoid $T^*G$), instead of the associated momentum map $M\raw\g^*$. This refinement is, of course, essential when $G$ is discrete.} Hence the quantization of the dual pair $pt\law M\raw \g^*$ should be an element of
the Kasparov group $KK(\C,C^*(G))$.

 This element can be defined when the $G$-action is proper and cocompact (i.e.\ $M/G$ is compact), and lifts to an action on the principal bundle $P$ defining the $\spinc$ structure. Namely, in that case one regards $\DS$ as an operator on the graded \Hs\ $L^2(M,\CS_n)$ of $L^2$-sections of $\CS_n$, which at the same time carries a natural \rep\ $\pi$ of $C_0(M)$ by multiplication operators, as well as a natural unitary \rep\ $U(G)$. 
 Provided that in addition the Dirac operator $\DS$ is almost $G$-invariant in the sense that $[U(x),\DS]$ is bounded for each $x\in G$, 
 these data specify an element $[L^2(M,\CS_n),\pi(C_0(M)),U(G),\DS]$ of 
the equivariant analytic K-homology group $K_0^G(M)=KK^G(C_0(M),\C)$ \cite{HR}. Here we suppress the grading of the \Hs\ in question in our notation. Let $$\mathrm{Index}_G:K_0^G(M)\raw K_0(C^*(G))$$ be the analytic assembly map as defined by  Baum, Connes, and Higson \cite{BCH}, seen however as a map taking values in $K_0(C^*(G))$ instead of $K_0(C_r^*(G))$ (cf.\ \cite{Val2} for this point). For simplicity we write
\beq
\mathrm{Index}_G(\DS_+)\equiv \mathrm{Index}_G([L^2(M,\CS_n),\pi(C_0(M)),U(G),\DS]).\eeq
 We then define the quantization of the dual pair $pt\law M\raw \g^*$ as
\beq
\CQ(pt\law M\raw \g^*)= \al_{C^*(G)}(\mathrm{Index}_G(\DS_+)), \label{Qpt} \eeq
where $\al_A: K_0(A)\raw KK(\C,A)$ is the natural isomorphism one has for
any  separable \ca\ $A$  \cite{Bla}.
As required, \er{Qpt} defines an element of $$KK(\CQ(pt),\CQ(\g^*))=KK(\C,C^*(G)).$$

For a much simpler example, whose significance will become clear shortly, 
consider the dual pair  $\g_-^*\hookleftarrow 0\raw pt$, where $0$ (seen as a coadjoint orbit of $G$) is the zero element of the vector space $\g^*$, equipped with minus the Lie--Poisson structure. 
Its quantization should be an element of the Kasparov \rep\ ring $KK(C^*(G),\C)$, which we simply take to be the graded \Hs\ $\H=\C\oplus 0$
carrying the trivial \rep\ of $G$, with $F=0$. We denote this element by 
$[\C,0,0]$, so that 
\begin{equation}
\CQ(\g^*\hookleftarrow 0\raw pt)=[\C,0,0]. \label{Q2}
\end{equation}
Let $$\ta_*: KK(\C,C^*(G))\raw KK(\C,\C)\cong\Z$$ be the map functorially induced by
the morphism $\ta:C^*(G)\raw \C$ given by the trivial \rep\ of $G$.\footnote{For $f\in C_c(G)$ one simple has $\ta(f)=\int_G dx\, f(x)$. This is the reason why we use $C^*(G)$ rather than $C^*_r(G)$, as is customary in the Baum--Connes conjecture: for
$\ta$ is not continuous on $C^*_r(G)$ (unless $G$ is amenable).}
 Functoriality of  the Kasparov product $$KK(\C,C^*(G))\x KK(C^*(G)),\C)\raw KK(\C,\C)\stackrel{\raw}{\cong}\Z$$ then  yields
\beq y\x [\C,0,0]=\ta_*(y) \eeq for any $y\in KK(\C,C^*(G))$.
In particular, \er{Qpt} and \er{Q2} give
\beq \CQ(pt\law M\raw \g^*)\x \CQ(\g^*_-\hookleftarrow 0\raw pt)=\ta_*(
\mathrm{Index}_G(\DS^M_+)); \label{FQq}\eeq
to avoid confusion later on, we have added a suffix $M$ to the pertinent Dirac operator. 

On the classical side, in the category $\GP$ we compute
\beq (pt\law M\raw \g^*) \circ (\g^*_-\hookleftarrow 0\raw pt)=pt\law M/\hspace{-1mm}/G\raw pt, \label{FQc} \eeq
where $M/\hspace{-1mm}/G$ is the Marsden--Weinstein quotient. Assuming that $M/\hspace{-1mm}/G$ is prequantizable (this is a theorem in the compact case \cite{GGK}), we have already seen from \er{Bott1}  that
\beq \CQ(pt\law M/\hspace{-1mm}/G\raw pt)=\mathrm{Index}(\DS^{M/\hspace{-1mm}/G}_+),\eeq
where we have denoted the appropriate Dirac operator on $M/\hspace{-1mm}/G$ by $\DS^{M/\hspace{-1mm}/G}$. 

Functoriality of quantization would imply
\beq \CQ(pt\law M\raw \g^*)\x \CQ(\g^*_-\hookleftarrow 0\raw pt)=
\CQ((pt\law M\raw \g^*) \circ (\g^*_-\hookleftarrow 0\raw pt)).\eeq
Using \er{FQq} and \er{FQc}, this amounts to
\beq \ta_*(\mathrm{Index}_G(\DS^M_+))=\mathrm{Index}(\DS^{M/\hspace{-1mm}/G}_+).\label{GGSC}\eeq

For $G$ and $M$ compact, this is precisely the so-called Guillemin--Sternberg conjecture that ``quantization commutes with reduction" \cite{GS} in its modern 
form \cite{GGK,Meinrenken,Sja}.\footnote{This conjecture is, in fact, a theorem \cite{JK,Meinrenken,Par,TZ}, but the name ``conjecture'' is still generally used.} To see this, note that for $M$ compact the Dirac operator $\DS_+$ is Fredholm, whereas for $G$ compact one has $K_0(C^*(G))\cong R(G)$, the \rep\ ring of $G$.
Consequently, $\mathrm{Index}_G(\DS_+)$ defines an element of $R(G)$,
and the map $\ta_*:R(G)\raw R(e)\cong \Z$ is just $[V]-[W]\mapsto \dim(V_0)-\dim(W_0)$, where $V_0\subset V$ is the space of $G$-invariant vectors, etc.

For $G$ countable (acting properly and cocompactly on $M$, as stated before), \er{GGSC}
boils down to  the naturality of the Baum--Connes assembly map for countable discrete groups \cite{Val2}. 
Combining  this fact with the validity of \er{GGSC} for compact $G$ and $M$,
 it can be shown that  \er{GGSC} holds for any  strongly Hamiltonian proper cocompact action of $G$ on a possibly noncompact symplectic manifold, provided that $G$ contains a discrete normal subgroup $\Gm$ with $G/\Gm$ compact \cite{HL}. 

Let us close this paper in the right groupoid  spirit by pointing out that all arguments in this section should be carried out for Lie groupoids instead of Lie groups. For example,  the pertinent  symplectic reduction procedure (generalizing Marsden--Weinstein reduction)
was first studied in \cite{MiW}, and can be reinterpreted in terms of the product in the category $\GP$ just as in the group case.
  A very interesting special case  comes from foliation theory, as follows (cf.\ \cite{C82,Con,CS,HS}). Let $(V_i,F_i)$, $i=1,2$, be foliations with associated holonomy groupoids $G(V_i,F_i)$ (assumed to be Hausdorff for simplicity). A smooth generalized
map $f$ between the leaf spaces $V_1/F_1$ and $V_2/F_2$ is defined as a 
principal bibundle $M_f$ between the Lie groupoids $G(V_1,F_1)$ and $G(V_2,F_2)$. Classically, such a bibundle defines a dual pair $T^*F_1\law T^*M_f\raw T^*F_2$ \cite{Lan4}. 
Here  $TF_i\subset TV_i$ is the tangent bundle to the foliation $(V_i,F_i)$, whose dual bundle $T^*F_i$ has a canonical Poisson structure.\footnote{The best way to see this is to 
interpret $TF_i$ as the Lie algebroid of $G(V_i,F_i)$.}   Quantum mechanically, $f$ defines an element \cite{C82,HS}
$$ f_!\in KK(C^*(G(V_1,F_1)), C^*(G(V_2,F_2))).$$
In the functorial approach to quantization, $f_!$ is interpreted as the quantization of the dual pair $T^*F_1\law T^*M_f\raw T^*F_2$. The functoriality of quantization among  dual pairs of the same type  should then follow from the computations in \cite{HS} on the quantum side and \cite{Lan4} on the classical side.  The construction and functoriality of shriek maps in \cite{AS1,C82} is a special case of this, in which the $V_i$ are both trivially foliated. 


\newpage
 \begin{thebibliography}{00}
\bibitem{AM} Abraham, R.; Marsden, J.E.: Foundations of mechanics. Second edition.
Benjamin/Cummings, Reading,  Mass.,  1978.
\bibitem{AE} Ali, S.T.; Englis, M.:  Quantization methods: A guide for physicists and analysts.  \texttt{arXiv:math-ph/0405065}.
\bibitem{AR} Anantharaman-Delaroche, C.; Renault, J.: Amenable groupoids.  Monographies de L'Enseignement Math\'{e}matique 36. L'Enseignement Math\'{e}matique, Geneva,  2000.
\bibitem{AS1} ÊAtiyah, M.F.; Singer, I.M.: The index of elliptic operators. I.  Ann. of Math. (2)  87  (1968), 484--530.
\bibitem{BCH} ÊBaum, P.; Connes,  A.; Higson, N.: Classifying space for proper actions and $K$-theory of group $C\sp  *$-algebras.  $C\sp *$-algebras: 1943--1993,   240--291, Contemp. Math., 167, Amer. Math. Soc., Providence, RI,  1994. 
\bibitem{Bla} Blackadar, B.: $K$-theory for operator algebras. Second edition.  Cambridge University Press, Cambridge,  1998.
\bibitem{Bott57} Bott, R.: Homogeneous vector bundles.  Ann. of Math. (2)  66  (1957), 203--248.
\bibitem{BR2} Bratteli, O.; Robinson, D.W.: Operator algebras and quantum statistical mechanics. 1. $C\sp *$- and $W\sp *$-algebras, symmetry groups, decomposition  of states. Second edition. Springer-Verlag, New York,  1987.
\bibitem{Cad}   Cadet, F.: D\'{e}formation et quantification par groupo\"{\i}de des vari\'{e}t\'{e}s toriques.  Ph.D.\ thesis,
Universit\'{e} d'Orl\'{e}ans, 2001.
\bibitem{Weinsteinplus} Cannas da Silva, A.; Weinstein, A.: Geometric models for noncommutative algebras. Berkeley Mathematics Lecture Notes, 10. American Mathematical Society, Providence, 1999.
\bibitem{C82} Connes, A.:  A survey of foliations and operator algebras.  
  Operator algebras and applications, Part I,   pp. 521--628,   Proc. Sympos. Pure Math., 38,  
 Amer. Math. Soc., Providence, R.I.,  1982. 
\bibitem{Con} Connes, A.: Noncommutative geometry. Academic Press, Inc., San Diego, CA,  1994.
\bibitem{Con2000} Connes, A.: Noncommutative geometry year 2000.  Highlights of mathematical physics (London, 2000),   49--110, Amer. Math. Soc., Providence, RI,  2002. 
\bibitem{ConnesLandi} Connes, A.; Landi, G.: 
  Noncommutative manifolds, the instanton algebra and isospectral  deformations.   Comm. Math. Phys.  221  (2001), 141--159.
\bibitem{CS}   Connes, A.; Skandalis, G.: 
The longitudinal index theorem for foliations. Publ. Res. Inst.
Math.\ Sci. 20  (1984)  1139--1183. 
\bibitem{CDW} Coste, A.; Dazord, P.; Weinstein, A.:  Groupo\"{\i}des symplectiques.  
  Publications du D\'{e}partement de Math\'{e}matiques. Nouvelle S\'{e}rie. A,  Vol. 2,   i--ii, 1--62,  Publ. D\'{e}p. Math. Nouvelle S\'{e}r. A, 87-2,  
 Univ. Claude-Bernard, Lyon,  1987. 
\bibitem{Cou} ÊCourant, T.J.: Dirac manifolds.  Trans. Amer. Math. Soc.  319  (1990),  631--661.
\bibitem{CF1} Crainic, M.; Fernandes, R.L.:
  Integrability of Lie brackets.  
  Ann. of Math. (2)  157  (2003),  575--620.
\bibitem{CF2}  Crainic, M.; Fernandes, R.L.: Integrability of Poisson brackets.  \texttt{arXiv:math.DG/0210152}.
\bibitem{Dir64} Dirac, P.A.M.: Lectures on quantum
mechanics. Belfer School of Science, Yeshiva University, New
York, 1964.
\bibitem{DKR} Doplicher, S; Kastler, D.; Robinson, D.W.: Covariance algebras in field theory and statistical mechanics.  Comm. Math. Phys.  3  (1966), 1--28.
\bibitem{EKQR}  Echterhoff, S.;  Kaliszewski, S.; Quigg, J.;  Raeburn, I.:
A Categorical Approach to Imprimitivity Theorems for C*-Dynamical   Systems.    \texttt{arXiv:math.OA/0205322}.
\bibitem{EH} Effros, E.G.; Hahn, F.: Locally compact transformation groups and $C\sp{*} $-  algebras. Memoirs of the American Mathematical Society, No. 75 American Mathematical Society, Providence, R.I.  1967
\bibitem{Emch} ÊEmch, G.G.: Mathematical and conceptual foundations of 20th-century physics. Amsterdam: North-Holland, 1984.
\bibitem{Faith73} Faith, Carl:  Algebra: rings, modules and categories. I.    
 Springer-Verlag, New York-Heidelberg,  1973.
\bibitem{Gli1}ÊGlimm, J.: Locally compact transformation groups.  Trans. Amer. Math. Soc.  101  (1961), 124--138.
\bibitem{Gli2} Glimm, J.: 
  Families of induced representations.  
  Pacific J. Math.  12  (1962), 885--911.
  \bibitem{Got1} Gotay, M.J.: Functorial geometric quantization and Van Hove's theorem.  Internat. J. Theoret. Phys.  19  (1980),  139--161.
 \bibitem{Got2} Gotay, M.J.: On the Groenewold-Van Hove problem for $\bold R\sp {2n}$.  J. Math. Phys.  40  (1999),   2107--2116. 
\bibitem{Green} Green, P.: The local structure of twisted covariance algebras.  Acta Math.  140  (1978),  191--250. 
\bibitem{GS} Guillemin, V.; Sternberg, S.: Geometric quantization and multiplicities of group  representations.  Invent. Math.  67  (1982),  515--538.
\bibitem{GSbook} Guillemin, Victor; Sternberg, Shlomo:
  Symplectic techniques in physics.  
 Cambridge University Press, Cambridge,  1984. 
\bibitem{GGK} Guillemin, V.; Ginzburg, V.; Karshon, Y.:
  Moment maps, cobordisms, and Hamiltonian group actions.   American Mathematical Society, Providence, RI,  2002.
\bibitem{Haa} Haag, R.:  Local quantum physics. Fields, particles, algebras. Second edition. Springer-Verlag, Berlin,  1996.
\bibitem{Hae} Haefliger, A.: Groupo\"{\i}des d'holonomie et classifiants.   Ast\'{e}risque  116  (1984), 70--97.
\bibitem{Heis}  Heisenberg, W.:
\"{U}ber die quantentheoretische Umdeutung kinematischer und mechanischer Beziehungen, Z. Phys. 33 (1925) 879-893. English translation in {\it Sources of \qm}, ed. B.L. van der Waerden (North-Holland, Amsterdam, 1967). 
\bibitem{HR} Higson, N.; Roe, J.:
  Analytic $K$-homology.  
 Oxford University Press, Oxford,  2000. 
\bibitem{HS} Hilsum, M.; Skandalis, G.: Morphismes $K$-orient\'{e}s d'espaces de feuilles et fonctorialit\'{e} en  th\'{e}orie de Kasparov (d'apr\`{e}s une conjecture d'A. Connes). Ann. Sci. \'{E}cole Norm. Sup. (4)  20  (1987),  325--390.
\bibitem{HL}  Hochs, P.;  Landsman, N.P.: The Guillemin--Sternberg conjecture for noncompact groups and spaces. To appear. 
\bibitem{Hueb} Huebschmann, J.: Lie-Rinehart algebras, Gerstenhaber algebras and Batalin-Vilkovisky  algebras.  Ann. Inst. Fourier (Grenoble)  48  (1998),   425--440.
\bibitem{JK} Jeffrey, L.C.; Kirwan, F.C.: Localization and the quantization conjecture.  Topology  36  (1997), 647--693.
\bibitem{JoMo} J\o rgensen, P.E.T.; Moore, R.T.: Operator commutation relations. Dordrecht: Reidel, 1984.
\bibitem{Kar} Karasev, M.V.:
  Analogues of objects of the theory of Lie groups for nonlinear Poisson  brackets. 
  Izv. Akad. Nauk SSSR Ser. Mat.  50  (1986),   508--538, 638.
\bibitem{Kar89} Karasev, M.V.: The Maslov quantization conditions in higher cohomology and analogs of  notions developed in Lie theory for canonical fibre bundles of symplectic  manifolds. I, II.  Selecta Math. Soviet.  8  (1989),   213--234, 235--258. 
\bibitem{KM} Krishnaprasad, P.S.; Marsden, J.E.: Hamiltonian structures and stability for rigid bodies with flexible  attachments.  Arch. Rational Mech. Anal.  98  (1987),   71--93. 
\bibitem{Lan} ÊLandsman, N.P.: Mathematical topics between classical and quantum mechanics. Springer Monographs in Mathematics. Springer-Verlag, New York,  1998.
\bibitem{Lan1} Landsman, N.P.: Lie groupoid $C\sp *$-algebras and Weyl quantization.  Comm. Math. Phys.  206  (1999),   367--381.
\bibitem{Lan2} Landsman, N.P.: Operator algebras and Poisson manifolds associated to groupoids.  Comm. Math. Phys.  222  (2001),   97--116. 
\bibitem{Lan3} Landsman, N.P.: Quantized reduction as a tensor product.  Quantization of singular symplectic quotients,   137--180, Progr. Math., 198, Birkh\"{a}user, Basel,  2001.  
\bibitem{Lan4} Landsman, N.P.: The Muhly-Renault-Williams theorem for Lie groupoids and its classical  counterpart.  Lett. Math. Phys.  54  (2000),  43--59. 
\bibitem{Lan5} ÊLandsman, N.P.: Deformation quantization and the Baum-Connes conjecture.  Comm. Math. Phys.  237  (2003),  87--103.
  \bibitem{Lan6} Landsman, N.P.:  Quantization as a functor.  Quantization, Poisson brackets and beyond,   9--24, Contemp. Math., 315, Amer. Math. Soc., Providence, RI,  2002. 
 \bibitem{Lan7} Landsman, N.P.: Functorial quantization and the Guillemin--Sternberg conjecture. Twenty Years of Bialowieza: A Mathematical Anthology.
Eds: S.T. Ali et al., pp.\ 23--45. World Scientific, New Jersey, 2005.
\texttt{arXiv:math-ph/0307059}.
 \bibitem{Lan8} Landsman, N.P.:  Quantum mechanics and representation theory: the new synthesis.  Acta Appl. Math.  81  (2004),  167--189.
\bibitem{LPS}  Quantization of singular symplectic quotients. Edited by N.P. Landsman, M. Pflaum and M. Schlichenmaier. Progress in Mathematics, 198. Birkh\"{a}user Verlag, Basel,  2001.
\bibitem{LR} ÊLandsman, N.P.; Ramazan, B.:  Quantization of Poisson algebras associated to Lie algebroids.  Groupoids in analysis, geometry, and physics,   159--192, Contemp. Math., 282, Amer. Math. Soc., Providence, RI,  2001. 
\bibitem{Mackenzie95} Mackenzie, K.C.H.:
  Lie algebroids and Lie pseudoalgebras. 
  Bull. London Math. Soc.  27  (1995),   97--147.
\bibitem{McK} Mackenzie, K.C.H.: An introduction to \Lg s and \La s.
Cambridge University Press, Cambridge, 2005. 
\bibitem{Mac52} Mackey, G.W.:
  Induced representations of locally compact groups. I.  
  Ann. of Math. (2)  55  (1952), 101--139.
\bibitem{Mac68} ÊMackey, G.W.: Induced representations of groups and quantum mechanics. W. A. Benjamin, Inc., New York-Amsterdam; Editore Boringhieri,  Turin,  1968. 
\bibitem{Mac92} Mackey, G.W.: The scope and history of commutative and noncommutative harmonic  analysis. History of Mathematics, 5. American Mathematical Society, Providence, RI; London Mathematical  Society, London,  1992. 
\bibitem{Mac98} Mackey, G.W.: The relationship between
classical mechanics and quantum mechanics. Contemp. Math.
214 (1998), 91--110.
\bibitem{MR} Marsden, J.E.; Ratiu, T.S.: Introduction to mechanics and symmetry. 17. Springer-Verlag, New York,  1999.
\bibitem{MaRaWe} Marsden, J.E.; Ra\c tiu, T.; Weinstein, A.: Semidirect products and reduction in mechanics.  Trans. Amer. Math. Soc.  281  (1984),   147--177.
\bibitem{Meinrenken} Meinrenken, E.; Sjamaar, R.: Singular reduction and quantization.  Topology  38  (1999),   699--762. 
\bibitem{MiW} ÊMikami, K.; Weinstein, A.: Moments and reduction for symplectic groupoids.  Publ. Res. Inst. Math. Sci.  24  (1988),   121--140. 
\bibitem{Moe} Moerdijk, I.: Toposes and groupoids.  Categorical algebra and its applications,   280--298, Lecture Notes in Math., 1348, Springer, Berlin,  1988.
\bibitem{MM} Moerdijk, I.; Mr\v cun, J.: Introduction to foliations and Lie groupoids. Cambridge Studies in Advanced Mathematics, 91. Cambridge University Press, Cambridge,  2003.
\bibitem{MRW} ÊMuhly, P.S.; Renault, J.N.; Williams,  D.P.: Equivalence and isomorphism for groupoid $C\sp *$-algebras.  J. Operator Theory  17  (1987),  3--22.
\bibitem{NT} ÊNest, R.; Tsygan, B.: Deformations of symplectic Lie algebroids, deformations of holomorphic  symplectic structures, and index theorems.  Asian J. Math.  5  (2001), 599--635.
 \bibitem{vNu}
 Neumann, J.\ von: Die Eindeutigkeit der Schr\"{o}dingerschen
Operatoren. Math. Ann. 104 (1931), 570--578.
\bibitem{JvN} Neumann, J. von:   Mathematische Grundlagen der
Quantenmechanik. Springer, Heidelberg, 1932.
\bibitem{NWX} Nistor, V.; Weinstein, A.; Xu, P.: 
  Pseudodifferential operators on differential groupoids.  
    Pacific J. Math.  189  (1999),  117--152.
\bibitem{Orsted79} \O rsted, B.: Induced representations and a new proof of the imprimitivity theorem. J. Funct. Anal. 31 (1979), 
355--359.
\bibitem{Par}   Paradan, P.-E.: $\spinc$ quantization and the K-multiplicities of the discrete series. \texttt{arXiv:math.DG/0103222}.
\bibitem{Pat} Paterson, A.L.T.: Groupoids, inverse semigroups, and their operator algebras.  Birkh\"{a}user Boston, Inc., Boston, MA,  1999.
\bibitem{Pauli} Pauli, W.:  \"{U}ber den Einflu\ss\ der Geschwindigkeitsabh\"{a}ngigkeit der Elektronenmasse auf den Zeemaneffekt.   Z. Phys. 31 (1925), 373--385.
\bibitem{Ped1}  Pedersen, G.K.: \ca s and their automorphism groups.
 Academic Press, London, 1979.
\bibitem{Ped2} Pedersen, G.K.: Analysis now.  Springer, New
York, 1989.
\bibitem{Phi} Phillips, J.:
  The holonomic imperative and the homotopy groupoid of a foliated  manifold.  
  Rocky Mountain J. Math.  17  (1987),  151--165.
\bibitem{Pri} Primas, H.: Chemistry, quantum mechanics and reductionism. Second Edition. Berlin: Springer, 1983.
\bibitem{Raca} Racani\`{e}re, S. Quantisation of Lie--Poisson manifolds. \texttt{arXiv:math.DG/0411066}. 
\bibitem{Ram}   Ramazan, B.:  
Deformation quantization of Lie--Poisson manifolds, Ph.D.\ thesis,
Universit\'{e} d'Orl\'{e}ans, 1998.
\bibitem{RR} Groupoids in analysis, geometry, and physics.  
 Edited by A. Ramsay and J. Renault. Contemporary Mathematics, 282.  
 American Mathematical Society, Providence, RI,  2001.
\bibitem{Ren} Renault, J.: A groupoid approach to $C\sp{*} $-algebras. Lecture Notes in Mathematics, 793. Springer, Berlin,  1980.
\bibitem{Rie72} Rieffel, M.A.: On the uniqueness of the Heisenberg commutation relations.  Duke Math. J.  39  (1972), 745--752. 
\bibitem{Rie74}  Rieffel, M.A.:  Induced representations of $C\sp{*} $-algebras.  Advances in Math.  13  (1974), 176--257. 
\bibitem{Rie82} Rieffel, M.A.: Applications of strong Morita equivalence to transformation group  $C\sp{*} $-algebras.  Operator algebras and applications, Part I,   pp. 299--310, Proc. Sympos. Pure Math., 38, Amer. Math. Soc., Providence, R.I.,  1982.
\bibitem{Rie89} Rieffel, M.A.: Deformation quantization of Heisenberg manifolds.  Comm. Math. Phys.  122  (1989),   531--562. 
\bibitem{Rie94} Rieffel, M.A.:
  Quantization and $C\sp *$-algebras.   $C\sp *$-algebras: 1943--1993,   66--97,  
 Contemp. Math., 167,  
 Amer. Math. Soc., Providence, RI,  1994. 
\bibitem{Schmuedgen} Schm\"{u}dgen, K.: Unbounded operator algebras and representation theory. Birkh\"{a}user Verlag, Basel,  1990. 
\bibitem{Wave} Schr\"{o}dinger, E.: Quantisierung als Eiegnwertproblem.
Ann. d. Physik 79 (1926), 361--376, 489--527.
\bibitem{Schweizer} ÊSchweizer, J.: Crossed products by $C\sp *$-correspondences and Cuntz-Pimsner  algebras.  $C\sp *$-algebras (M\"{u}nster, 1999),   203--226, Springer, Berlin,  2000.
\bibitem{Sja} Sjamaar, R.: Symplectic reduction and Riemann-Roch formulas for multiplicities.  Bull. Amer. Math. Soc. (N.S.)  33  (1996),  327--338.
\bibitem{StadlerUchi} Macho Stadler, M.; O'uchi, M.: Correspondence of groupoid $C\sp *$-algebras.  J. Operator Theory  42  (1999),   103--119.
\bibitem{Thi} Thirring, W. (1983).   A course in mathematical physics. Vol.\ 4:
Quantum mechanics of large systems. New York: Springer-Verlag.
\bibitem{TZ} Tian, Y.; Zhang, W.: An analytic proof of the geometric quantization conjecture of  Guillemin-Sternberg.  Invent. Math.  132  (1998),   229--259.
\bibitem{Val2} Mislin, G.; Valette, A.: Proper group actions and the Baum--Connes conjecture. BirkhŠuser Verlag, Basel,  2003.
  \bibitem{vanH}  van Hove, L.: Sur certaines representations unitaires d'un groupe infini de transformations, Mem. Acad. Roy. Belg. 26 (1951), 61-102.
 \bibitem{Wei87} Weinstein, A.: Symplectic groupoids and Poisson manifolds.  Bull. Amer. Math. Soc. (N.S.) 16 (1987), 101--104.
\bibitem{Wei83} Weinstein, A.: The local structure of Poisson manifolds.  J. Differential Geom.  18  (1983),   523--557.
\bibitem{Wei89}  Weinstein, A.: Blowing up realizations of
Heisenberg-Poisson manifolds.    Bull.\ Sc.\ math. (2) 113,  (1989)
381-406.
\bibitem{WeinsteinXu} Weinstein, A.; Xu, P.: Extensions of symplectic groupoids and quantization.  J. Reine Angew. Math.  417  (1991), 159--189. 
 \bibitem{Wei2} Weinstein, A.:
  Groupoids: unifying internal and external symmetry. A tour through some  examples.    Groupoids in analysis, geometry, and physics, 1--19,  
 Contemp. Math., 282,   Amer. Math. Soc., Providence, RI,  2001. 
 \bibitem{Xu1} ÊXu, P.: Morita equivalence of Poisson manifolds.  Comm. Math. Phys.  142  (1991),   493--509. 
\bibitem{Xu2} Xu, P.: Morita equivalent symplectic groupoids.  Symplectic geometry, groupoids, and integrable systems,   291--311, Math. Sci. Res. Inst. Publ., 20, Springer, New York,  1991. 
\bibitem{Zak} Zakrzewski, S.: Quantum and classical pseudogroups. II. Differential and symplectic  pseudogroups.  Comm. Math. Phys.  134  (1990),  371--395.
\bibitem{Ziegler} Ziegler, F.:  M\'{e}thode
 des orbites et repr\'{e}sentations quantiques. Ph.D.\ thesis,
 Universit\'{e} de Provence, 1996.
\end{thebibliography}
 \end{document}